\documentclass[useAMS, usenatbib]{mnras}
\usepackage{aas_macros}
\usepackage{amsmath}
\usepackage{amssymb} 
\usepackage{graphicx} 
\usepackage{float} 
\usepackage{amsfonts}
\usepackage{lmodern}
\usepackage{bm}
\usepackage{placeins}
\usepackage{color}
\usepackage{orcidlink}
\usepackage{ulem}

\defcitealias{Smith2020}{Paper I}

\newcommand{\bc}{\begin{center}}
\newcommand{\ec}{\end{center}}
\newcommand{\Msun}{\mathrm{M_\odot}}
\newcommand{\Zsun}{\mathrm{Z_\odot}}
\newcommand{\mpart}{m_\mathrm{part}}

\newcommand{\simSN}{\textit{SN}}
\newcommand{\simPI}{\textit{PI}}
\newcommand{\simPE}{\textit{PE}}
\newcommand{\simSNPIPE}{\textit{SN-PI-PE}}
\newcommand{\simNoFB}{\textit{NoFB}}

\let\oldhat\hat
\renewcommand{\hat}[1]{\oldhat{\mathbf{#1}}}
\let\oldbullet\bullet \renewcommand{\bullet}[1][0pt]{ 
\mathrel{\raisebox{#1}{$\oldbullet$}}
}
\title[IMF averaging versus IMF sampling]{The sensitivity of stellar feedback to IMF averaging versus IMF sampling in galaxy formation simulations}
\author[M. C. Smith]{
Matthew C. Smith\orcidlink{0000-0002-9849-877X}$^{1}$\thanks{E-mail: matthew.smith@cfa.harvard.edu}\\
  $^1$ Harvard-Smithsonian Center for Astrophysics, 60 Garden Street, Cambridge, MA 02138, USA \\
  }

\begin{document}

\maketitle

\begin{abstract}
Galaxy formation simulations frequently use Initial Mass Function (IMF) averaged feedback prescriptions, where star particles are assumed to represent single stellar populations that fully sample the IMF. This approximation breaks down at high mass resolution, where stochastic variations in stellar populations become important. We discuss various schemes to populate star particles with stellar masses explicitly sampled from the IMF. We use Monte Carlo numerical experiments to examine the ability of the schemes to reproduce an input IMF in an unbiased manner while conserving mass. We present our preferred scheme which can easily be added to pre-existing star formation prescriptions. We then carry out a series of high resolution isolated simulations of dwarf galaxies with supernovae, photoionization and photoelectric heating to compare the differences between using IMF averaged feedback and explicitly sampling the IMF. We find that if supernovae are the only form of feedback, triggering individual supernovae from IMF averaged rates gives identical results to IMF sampling. However, we find that photoionization is more effective at regulating star formation when IMF averaged rates are used, creating more, smaller \ion{H}{ii} regions than the rare, bright sources produced by IMF sampling. We note that the increased efficiency of the IMF averaged feedback versus IMF sampling is not necessarily a general trend and may be reversed depending on feedback channel, resolution and other details. However, IMF sampling is always the more physically motivated approach. We conservatively suggest that it should be used for star particles less massive than $\sim500\,\Msun$.
\end{abstract}


\section{Introduction} \label{Introduction}
Substantial progress in understanding the formation and evolution of galaxies has been made over the last few decades
by accounting for the role played by stellar feedback \citep[see e.g. the reviews of][and the references therein]{Somerville2015,Naab2017}.
Stars interact with the gas inside galaxies and beyond via a variety of complex processes, including supernovae (SNe), stellar winds and radiation
(providing photoionization, photodissociation, photoheating and radiation pressure). They can therefore act to influence the thermal and kinetic state of 
the interstellar medium (ISM), regulate star formation, enrich gas with metals, drive galactic outflows (carrying metals out into the circumgalactic medium (CGM)) and more besides. As the preferred terminology, ``feedback'', would suggest, these processes represent a
back-reaction on the gas from which stars themselves originate. Thus, numerical simulations of galaxies must account for the gaseous and
stellar components, as well as their interactions, in a self-consistent manner.

The ``star particle'' has been a nearly ubiquitous feature of hydrodynamic simulations of galaxy formation for nearly three decades
\cite[see early examples in e.g.][]{Katz1992a,Katz1996a,Navarro1993,Mihos1994}. In the majority of cases, the star particle does not
represent individual stars or even star clusters, per se. It instead traces an underlying collisionless stellar fluid
in much the same way that a dark matter particle traces the dark matter fluid. Unlike the dark matter component, however, the stellar
component must interact with the gas through processes in addition to gravity. Star particles can be created from gas mass to represent
the process of star formation. Stellar feedback can be directly tied to the star particle \citep[as in][one of the first
examples of stellar feedback being explicitly included in a simulation]{Katz1992a}. However, when the mass resolution is low, some or all of the stellar feedback
may not be directly associated with star particles but instead associated with the star forming gas. For example, \cite{Springel2003}
presents models which treat unresolved SN feedback by modifying the equation of state of star forming gas and 
launching stellar feedback driven
winds from the gas without directly involving a star particle. There is no inconsistency in
this context as the star particle does not explicitly represent the location of the stellar mass, which is in some sense smoothed over
the ensemble of particles. Such ``diffuse'' stellar feedback models are frequently 
used in modern large volume cosmological simulations, usually
in combination with star particle centred feedback to a greater or lesser extent 
\citep{Vogelsberger2013,Vogelsberger2014,Dubois2014,Dubois2016,Dave2016,Dave2019,Pillepich2018}.
Once the mass resolution of the star particle approaches $\sim10^4\,\Msun$ or better, stellar feedback is usually tied
directly to star particles and is modelled in a more explicit manner 
\citep[for a small sample of contemporary approaches see e.g.][]{Hopkins2014a,Hopkins2017a,Ceverino2014,Kimm2015,Agertz2015,Marinacci2019}.
This is appropriate because the lower particle masses begin to allow small scale spatial and temporal clustering of feedback processes
to be resolved. Regardless of the approach taken, a link between the mass in stars and the resulting stellar feedback budget must be made.
Because individual stars are not resolved in the schemes described above, this relationship is obtained by averaging over the statistical
distribution of stellar masses.

This distribution is given by the initial mass function (IMF). Combining an IMF with a set of stellar evolution models can produce
the net feedback properties (e.g. SN rate, luminosity in various bands relevant to feedback, wind power and mass loss rate etc.)
for a single stellar population (SSP) under the assumption that the IMF is fully sampled. Then, the feedback budget can
be determined for a star particle in a simulation (typically with a dependence on its age and metallicity via lookup tables), 
rescaled by its mass.
We will refer to this as IMF averaged feedback throughout this work. It is important to note that even when IMF averaged SN rates
are used, SNe can still be modelled as discrete events by stochastically sampling the rates 
\citep[see e.g.][]{Stinson2010,Hopkins2014a,Kimm2015,Revaz2016,Smith2018}.
Consistent evidence for a universal IMF has been amassed from observations
of large populations of stars, featuring a steep power law at high masses \citep{Salpeter1955} with a knee at $\sim1\,\Msun$
\citep{Kroupa1993,Kroupa2001,Chabrier2003}.
Potential deviations away from the universal IMF can arise in two forms. Firstly,
there has been some evidence that the IMF may vary systematically between galaxies 
\citep{Hoversten2008,vanDokkum2010,Cappellari2012,Conroy2012b,Kalirai2013,LaBarbera2013,Geha2013}, 
with possible correlations with central velocity
dispersion and/or metallicity. However, the magnitude of this phenomenon remains unclear. Nonetheless, if the IMF does indeed vary
systematically, this effect is conceptually simple to include in IMF averaged feedback schemes. For example, in
cosmological zoom-in simulations of Milky Way-like galaxies, 
\cite{Gutcke2019} use
an IMF that varies as a function of metallicity to adjust SN rates and metal enrichment from AGB winds on-the-fly.

Alternatively, deviations could arise due to variations of the IMF on small scales or as a result of undersampling a universal
IMF in small populations.
The total distribution of stars in a galaxy is a composite of the various individual star forming regions. Thus, an
integrated galactic IMF (IGIMF) arises from the combination of the IMF within individual star clusters and the cluster mass
function \citep{Kroupa2003}. If a common universal IMF is well traced within each cluster independent of cluster mass, then the
IGIMF will be identical to the universal IMF. If the IMF varies from cluster to cluster the shape of the IGIMF will change. In particular, if the high-mass cutoff is a function of cluster mass, then the IGIMF slope will be steeper than the universal IMF.
There is evidence that massive stars are rarer in low-SFR environments \citep{Meurer2009,Lee2009,Lee2016b,Gunawardhana2011}, which
could be a consequence of this scenario. It has been posited that this could arise due differences in how star formation proceeds in clusters of different mass, leading to an intrinsic relationship between cluster mass and high-mass cutoff 
\citep{Kroupa2003,Weidner2006,Weidner2010,Weidner2013}. However, it has also been argued that no such deterministic link exists and
that observations are consistent with a stochastically sampled uniform IMF
\citep{Elmegreen2006,Corbelli2009,Calzetti2010,Fumagalli2011,Andrews2013,Andrews2014}. These effects cannot be captured
a priori with IMF averaged feedback as they emerge as a result of stochastic effects arising from undersampling of the IMF in a given
small population of stars.

The alternative is to populate star particles at their birth with an inventory of stars by explicitly sampling the IMF. Feedback
budgets can then be based on the individual stars residing in the star particle. When star particles are massive enough, this will converge
with the IMF averaged approach because the IMF will be well sampled within the particle. Taking the opposite limit yields the modelling
of individual stars. The assumption that a stellar population samples the IMF well only holds for population masses above $\sim10^5\,\Msun$
\citep{Carigi2008,Revaz2016}, below which stochastic effects will begin to emerge. Explicit sampling of the IMF allows issues related
to undersampling of the IMF (as described in the previous paragraph) to be captured. 
It also allows the inhomogeneous distribution of stellar
feedback among stars (varying luminosities, mass loss, SN energy injection etc.) to be resolved. Various schemes for explicit
IMF sampling have been presented and used in simulations of individual GMCs, patches of discs (stratified boxes) and entire
galaxies 
\citep[see e.g.][some of which we will discuss in greater detail in Section~\ref{pop IMF}]{Gatto2017,Sormani2017,Geen2018,Hu2017,Hu2019,
Fujimoto2018,Emerick2019,Applebaum2020,Gutcke2021}. 

\cite{Su2018} estimate the effect of using stochastically sampled stellar masses compared to
IMF averaged feedback in cosmological zoom-in simulations. 
They find that as long as SNe are modelled as discrete events (even if sampled from IMF averaged rates),
stochastic variation of the stellar content of star particles does not produce a significant departure from simulations run with 
IMF averaged feedback. However,
they do not explicitly perform IMF sampling, instead using a toy model to modulate their IMF averaged feedback (OB winds,
luminosities and SN rates). In Section~\ref{Discussion} we will discuss the validity of such an approach. \cite{Grudic2019} use a similar
technique in simulations of individual GMCs, but find that approximating the inhomogeneous distribution of UV luminosities among
stars, leading to the presence of rare, bright sources, results in lower star formation efficiencies compared to using IMF averaged
feedback. In non-cosmological simulations of dwarfs, \cite{Applebaum2020} find that when SN feedback and H$_2$ dissociating radiation are
linked to explicitly sampled stellar masses rather than being based on IMF averaged rates (that still discretize SNe), 
feedback is moderately less
efficient at regulating SFRs and the mass of cold gas. We will discuss their findings in more detail in Section~\ref{Discussion}.

This work is laid out as follows. Section~\ref{pop IMF} contains an extended discussion of the details of explicit IMF sampling schemes.
We shall use some simple Monte Carlo numerical experiments to demonstrate the advantages and disadvantages of several schemes, populating
star particles of various masses. In Section~\ref{imfex vs imfav} we use high resolution non-cosmological simulations of dwarf 
galaxies to directly compare the use of IMF averaged feedback to explicit IMF sampling. In Section~\ref{Discussion} we
discuss our findings and their consequences in greater detail, as well as providing a comparison to some other relevant works. 
Section~\ref{conclusion}
presents our conclusions. Appendix~\ref{stoch} demonstrates that our simulations are robust to stochastic effects by rerunning
the early stages of a subset of our simulations with randomly perturbed initial conditions and different random number generator seeds.
\vspace{-4ex}
\section{Populating star particles from an IMF}\label{pop IMF}
\subsection{Requirements of an IMF sampling scheme}
One of the main reasons for paying close attention to details of the output IMF in a galaxy formation simulation is its
relationship to the feedback budget. Since the vast majority of stellar feedback arises from comparatively rare,
massive stars, subtle changes to the distribution of stellar masses can potentially influence the evolution of the
galaxy.
When designing a scheme to populate star particles created in a galaxy formation simulation with stellar
masses drawn from an IMF, there are two main requirements that are often in tension with each other. The scheme
must attempt to reproduce the input IMF as closely as possible, but it should also conserve stellar mass. Problems
arise because a sequence of discrete stellar masses drawn from an IMF is in general unlikely to sum exactly
to a previously specified value (e.g. the mass of a star particle), with the last sampled mass overshooting the target. What a
scheme does in this scenario determines how closely it prioritizes mass conservation vs. reproducing the IMF. 

One possible solution is to accept the last draw and source more mass to make up the difference. 
Such an approach is trivial to implement if the initial star particle mass is smaller than the mass of the gas resolution
element from which it is formed. Otherwise, the mass reservoir can be `topped up' by taking mass from other nearby gas 
particles/cells (either
by merging or partially draining them, see e.g. \citealt{Hirai2020,Gutcke2021}). 
If done in combination with the standard method for forming
star particles (stochastically sampling the SFRs of individual gas particles/cells independently to trigger the spawning
of or conversion to a star particle), care must be taken to avoid or minimize inconsistencies between the expected SFR of the gas and the rate at which
stellar mass is created (e.g. \citealt{Hu2019} avoids this issue by exchanging mass between star particles \textit{after}
they have been created). 
However, in this work we will avoid any additional transfer of mass to particles and focus on methods to populate a star
particle of fixed mass with stars, presenting a scheme that can be easily incorporated into any existing
implementation of star formation.

Before proceeding, we will comment on why it is important that an input IMF is accurately
reproduced in a galaxy formation simulation. In nature, the IMF is an emergent property of the process of star
formation, arising from the small scale physics governing the fragmentation and gravitational collapse
of turbulent, star forming gas and its interaction with
feedback. By
contrast, in galaxy formation simulations it is impossible to resolve the physical processes that give rise to the
IMF. An IMF must therefore be imposed upon the simulation as a sub-grid model, either in an explicit manner by sampling
from it on-the-fly or by taking IMF averaged approaches to stellar feedback. The chosen IMF may be derived
empirically from observations or it can be based on theoretical expectations. It can be a fixed, universal IMF or
it could take a more complicated form, being allowed to vary with galactic or even local properties. Regardless of
the form adopted, once chosen, the input IMF encodes the unresolved physics of small scale star formation that cannot
be captured in the simulation. Therefore, it is imperative that it is not subsequently biased by numerical 
issues originating from the implementation. As an obvious example, when the star particle mass resolution begins to approach (or
even drop below) the upper mass cut of the IMF, the resulting distribution of stellar masses is very vulnerable to being
biased away from input IMF (as we shall show below). Such a bias must be avoided as much as possible, since it is
obvious that the distribution should be independent of the mass resolution as this is entirely numerical. 

A more
subtle problem can occur in regions of low SFR. When a star particle is created and is
populated from the IMF, it is possible that there is not enough mass in the local star forming cloud (let alone
the star particle) to satisfy a draw from the input IMF. It is tempting under these circumstances to simply
discard the draw, since it is clear that such a massive star could not form in this environment. However, this also
leads to a biasing of the input IMF that has its basis in numerics not physics. If the resolved physics of the simulation
frequently provides situations where there is not enough gas mass available to accurately sample from the input IMF,
this suggests that the input IMF is inappropriate. A better IMF should be chosen that encodes either an empirical
or theoretical model for how star formation proceeds in low SFR environments. If draws are
discarded, then the resulting distribution of stellar masses is simply a flawed realisation of the input IMF
rather than reflecting any resolved physics. One could imagine that in the scenario described above, a different output
from the random number generator could have resulted in the massive star being formed later in the evolution of the
star forming cloud, when there was sufficient mass available. It is also possible that the manner in which the
cloud assembled is sensitive to resolution. The final distribution of stellar masses averaged over many star 
particles should have no such dependence on numerics since this introduces a bias with no physical basis. 
Additionally, we note that it is entirely possible that a low mass galaxy does not
form enough stars to fully populate the IMF. This does not necessarily indicate that the input IMF is inappropriate, rather that
multiple realisations of the simulation should be performed to assess the impact of stochasticity.

In the previous discussion, it is clear that the majority of cases in which the output IMF can be biased are due to
issues of mass conservation. While we have described why simply throwing away a draw from the IMF because of the lack
of stellar mass must be avoided, it is also apparent that simply accepting every draw would result in a
net overproduction of stars because the target mass will always be exceeded. 
An IMF sampling scheme must therefore balance the competing demands of mass conservation and
IMF preservation. The compromise usually takes the form of an inconsistency between the target mass and the mass of the
samples (which we will refer to as the assigned mass) on a particle by particle basis, but a consistency when averaged
over many particles. The particle level inconsistency may be eliminated after the fact by a transfer of mass (as
mentioned at the beginning of this section) or can simply be left as a mismatch between the dynamical and assigned mass
of the particle (the approach we adopt in this work). Because the exact N-body interactions
between stars cannot usually
be treated in a galaxy formation simulation (the use of softened gravitational forces often being adopted) this
inconsistency is not of much concern from a dynamical perspective.
The mass inconsistency can also be resolved conceptually if low mass stars (which do not
contribute substantially to the feedback budget) are not tracked explicitly. The mass inconsistency can then be
thought of as a simple redistribution of the low mass stars between star particles, as long as the overall
inconsistency sums to zero over many particles (e.g. \citealt{Applebaum2020} use this philosophy). This has no practical
impact on the operation of the scheme, other than to rationalise the discrepancy. Additionally, it 
obviously does not work in the case where the
combined mass of the explicitly tracked massive stars exceeds the dynamical mass.
Regardless, if the inconsistency is too large there may
not be sufficient mass to return via stellar winds and SN ejecta as required by the assigned stellar masses. We will
briefly touch on this subject later in this work.

Finally, we note that we have been vague as to distinguishing between a truly universal IMF versus an IGIMF arising
from an IMF that varies between clusters, as described in Section~\ref{Introduction}. 
For example, if stars born in clusters follow the shape of a universal IMF, 
but properties of the cluster (most importantly the mass) impose a high mass cutoff then the IGIMF (which is composed
of the sum of the IMFs of the individual clusters) will be steeper than the canonical IMF \citep{Kroupa2003}. This
means that the IGIMF will have a dependence on the mass function of the star forming regions with galaxy properties
(mass, metallicity etc.) playing a key role. However, it is important to note that star particles should not be 
conflated with physical star clusters. Unless the adopted star formation prescription produces a distribution
of star particle masses such that the mass function is an emergent property of the simulation (and is believed to be
an accurate representation of the true cluster mass function), the star
particle mass has no physical meaning. In almost all cases the star particle mass is set uniformly 
as a parameter of the simulation or as a consequence of the gas resolution.
This means that truncation of the upper end of a universal IMF based on the star particle mass is unphysical. If
the use of a particular form of a cluster mass dependent IMF (and resulting IGIMF) is required, 
a more complex sub-grid model is needed. 

\subsection{Methods for explicit IMF sampling} \label{pop IMF methods}
We will now explore several methods for populating star particles with stellar masses from the IMF. As explained in the previous paragraph, we 
do not consider methods that account for the truncation of the high mass end of the IMF as a function of the desired
total sample mass (the target mass) \citep[e.g. the `sorted sampling' method of][]{Weidner2006} as in general star particles do
not accurately represent stellar clusters. Three simple methods of sampling from the IMF to reach a target mass are
`stop before', `stop after' and `stop nearest' \citep[see ][]{Haas2010}. These terms refer to what is done with the 
last drawn stellar mass which will inevitably exceed the desired target. 

`Stop before' sampling discards the last drawn
stellar mass. It guarantees that the total drawn stellar mass (which we refer to as the assigned mass, $m_\mathrm{asn}$) will not
exceed the mass of the star particle. However, this means that the sum of assigned masses over all particles will
inevitably show a deficit relative to the total dynamical mass of star particles formed. It will also create a bias towards low mass
stars since a high mass star is more likely to exceed the target mass. The output IMF will therefore deviate downwards
from the input IMF, being finally truncated at the star particle mass. As the star particle mass is increased, this
bias is reduced. Likewise, the fractional deficit of the total assigned mass to the total particle mass will reduce because
the relative contribution of the last draw is reduced. `Stop after' sampling takes the opposite approach, always
keeping the last drawn star. By avoiding discarding any draw, this method guarantees that the
distribution of assigned stellar masses perfectly follows the shape of the input IMF. The downside is that this comes
at the cost of normalizing the IMF upwards, always assigning more stars to particles than the available stellar mass.
Again, the relative impact of this bias is reduced as the star particle mass is increased. It should also be noted that
any scheme which accepts the last drawn star and then makes the sampled and dynamical masses consistent by taking additional
mass from nearby gas \textit{after} the decision to create a particle has already been made will suffer a similar bias, 
artificially inflating the local SFR. 

A compromise between these two methods is the `stop nearest' approach. With this scheme, the last drawn stellar mass is
kept if the total assigned mass is closer to the target mass with its inclusion than without. The motivation is that
the cases of the last drawn mass being discarded will be counterbalanced by the occasions it is kept. This is true for large star particle masses, where the size of the gap between the penultimate drawn mass and the target mass is
insignificant compared to the target mass. However, as we shall demonstrate, this method suffers from the same biases
as the `stop before' scheme (albeit to a lesser extent) when the star particle mass is smaller.

\begin{figure}
\centering
\includegraphics{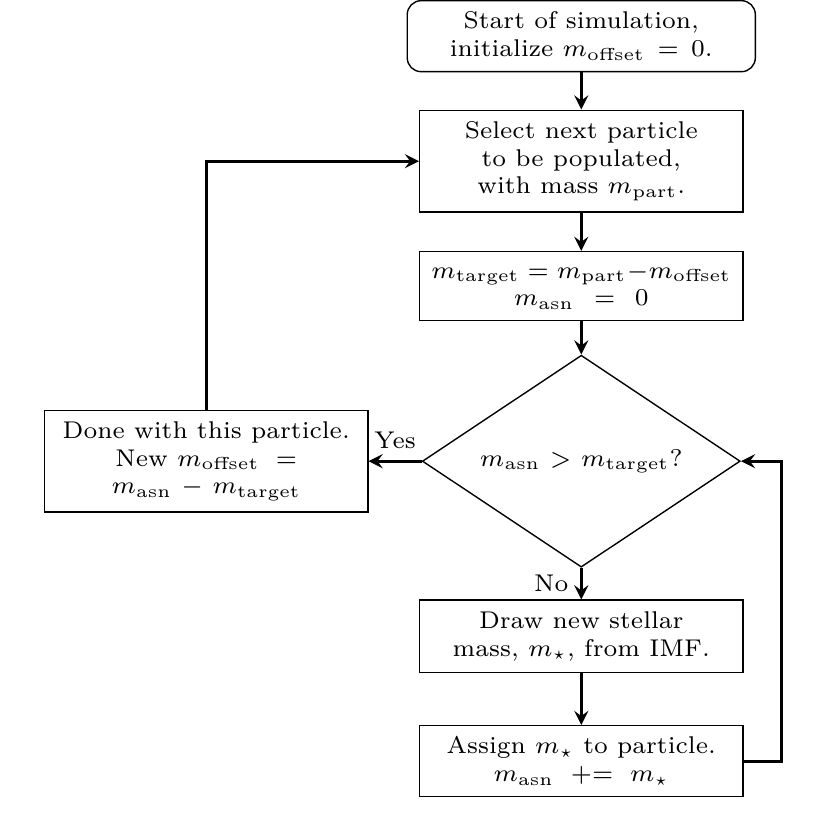}
\caption{A schematic illustration of our `adjusted target' star particle populating scheme. Stellar masses are drawn
from the input IMF until a target mass is exceeded. The last drawn mass is always kept, ensuring that
the output IMF is unbiased. The target mass is not in general equal to the dynamical mass of the star particle, but
compensates for the amount that the target was overshot when sampling the previous star particle (the offset mass). Thus,
some star particles will have more assigned stellar mass than their dynamical mass while the reverse is true for others.
Across an ensemble of star particles the total assigned
mass is equal to the total dynamical mass and the input IMF is perfectly reproduced.}
\vspace{-4ex}
\label{fig_flow} 
\end{figure}

A common deficiency of these three methods is that the populations of individual particles are completely independent of each
other. This is only appropriate if the star particle represents an independent single stellar population (SSP) (e.g. as
formed in a star cluster). It is
inappropriate for small star particle masses since the ensemble of star particles represent the stellar population
together. However, as discussed earlier, it is usually impossible at the resolutions of a galaxy formation simulation for 
a realistic population of stars to be an emergent feature of the local physical environment. Thus, in order to
accurately recover an input IMF, each star particle cannot be populated in isolation. Instead, information about
the distribution of stellar masses in previous star particles must be taken into account to correct for biases.
While this necessarily involves some form of `action at a distance', it is important to note that
this was already implied by requiring that ensembles of stars follow an enforced IMF together rather than
setting individual stellar masses independently from resolved local conditions.
We implement this approach in a scheme we refer to as `adjusted target'. Fig.~\ref{fig_flow} contains a
flowchart detailing the algorithm. The scheme is similar to `stop after' except that the target mass is not in 
general the same as the particle mass. Instead, the target mass for a given particle is adjusted based on 
how far the previous star particle overshot its target.
This allows us to keep the total assigned mass consistent
with the total stellar mass formed across many particles, but to simultaneously perfectly reproduce the input IMF because we never discard a draw. This scheme is very similar to that presented in \cite{Hu2017} with the difference that we do not adjust the dynamical mass
of particles to enforce consistency between assigned mass and dynamical mass for each particle individually. This avoids unphysical mass transfer over potentially arbitrary distances.\footnote{The scheme of \protect\cite{Hu2019} 
addresses this issue, but involves the additional complications of mass transfer between star particles.}
Similarly, there is no inter-particle mass exchange so it has the advantage
that it can be trivially implemented on top of any pre-existing implementation of star formation with a minimum of additional
coding.

When a star particle is populated, we draw stellar masses from the input IMF and assign them to the star particle
until the total assigned mass exceeds the target. The last draw is kept and the star particle is now finished with.
The `offset mass' is equal to the sum of masses assigned to the particle minus the target. 
The target mass of the next particle to be
populated is set to its particle mass minus the offset mass from the previous particle. The target mass 
will always be equal to or less than the particle mass. However, while the assigned
stellar mass will always exceed the target mass it will not always exceed the particle mass. Some particles will
have an excess of assigned stellar mass while others will have a deficit. Averaged across multiple star particles
the total assigned mass will equal the total star particle dynamical mass.

Without modification, the algorithm presented in Fig.~\ref{fig_flow} accounts for circumstances where the
target mass is negative or zero. This occurs if the last assigned stellar mass of the previous particle resulted in a substantial
overshoot of the target. When the target mass is negative or zero, the algorithm will draw no stellar masses for the star
particle and the offset mass is reduced by the mass of the star particle, resulting in a higher target mass for the next
star particle. This feature of the adjusted target scheme means that it is possible to include stellar masses that are
larger than the fiducial star particle mass, simply resulting in a particle that has more assigned mass than dynamical mass
and some particles that compensate by having no assigned mass. For example, consider the extreme example where a $100\,\Msun$
star is the first draw for a $20\,\Msun$ star particle and, for the sake of simplicity, the current offset mass is $0\,\Msun$ (meaning that the target
mass for this star particle is also $20\,\Msun$). The draw of the $100\,\Msun$ star is accepted and assigned to the star particle. The target mass has been exceeded, so sampling is now concluded for this star particle. The offset mass is now
$80\,\Msun$ because the target was overshot by that amount. The algorithm will produce target masses for the next four
$20\,\Msun$ particles that are less than or equal to zero. They will therefore have no assigned stellar mass. 
The resulting
total assigned mass equals the total dynamical mass across the five particles, at the cost that each particle has an
inconsistency between its individual assigned and dynamical masses. As mentioned above, because we cannot anyway resolve
the dynamics of individual stars this is largely irrelevant. It is only a problem if there is not enough mass available
to be returned through feedback (discussed below). In practice, cases such as this are rare because high mass stars are proportionally less likely to be drawn. However, as we shall
demonstrate, failing to allow for their formation biases the IMF and reduces the overall feedback budget. The target mass
usually stays close to the particle mass (unless the particle mass is considerably smaller than typical stellar mass), only
moving substantially if a very massive star is drawn.

As described here, our algorithm only ensures the input IMF is reproduced over the sequence
of sampled star particles. It does not account for spatial information, so it does not guarantee that a sub-volume of a 
simulation will contain a population of stars that reproduce the IMF. In principle, multiple concurrent 
realisations of the underlying ``book-keeping'' can be used. For example, in a
cosmological simulation it would make sense to define the ``next star particle'' as the next formed in the same galaxy,
maintaining separate offset masses for individual systems, ensuring that the IMF is maintained in each. For a
non-cosmological simulation the galaxy could likewise by divided into separate regions where the algorithm
is applied independently, maintaining the IMF locally.
However, in practice we find that this is unnecessary since it is
extremely unlikely that applying the algorithm globally results in, for example, 
all the massive stars being formed at one side of the galaxy.
A regional division may be necessary if an input IMF that has a dependence on
local conditions is used such that IMF varies strongly within a galaxy at a given moment in time. If the IMF varies slowly as a
function of time (compared to the time-scale on which $m_\mathrm{offset}$ fluctuates, $\sim m_\mathrm{part}/\dot{M}_\star$) and remains relatively uniform within a galaxy then this is not necessary. It should also be noted that our scheme as proposed does not permit
the spatial clustering of stars as a function of their mass to be directly specified, with correlations simply arising from completely random
sampling of the IMF, although this is unlikely to be of much concern at the resolution of global galaxy simulations.

Finally, we will briefly mention some other sampling methods that have been used in the literature that do not populate a star
particle with a sequence of drawn stellar masses. In stratified box simulations,
\cite{Gatto2017} populate sink particles with stellar masses drawn from the IMF. They only consider massive stars, sampling
once from a mass range of $9-120\,\Msun$ every time $120\,\Msun$ is accreted onto the sink and assume the remaining mass forms
lower mass stars. However, this scheme can produce an overall deficit of massive stars because sink particle masses are not in general
a multiple of $120\,\Msun$. \cite{Geen2018} adopt a similar approach in simulations of GMCs. \cite{Sormani2017} present a scheme which
instead divides the IMF into a number of bins and populates particles by performing Poisson sampling to determine the number of stars
in each bin based on the expected number. This has the advantage of reproducing the IMF in an unbiased, albeit discretized, 
form across an ensemble of particles, at the cost of a discrepancy between the sampled and dynamical mass within a given particle.
No additional correction is necessary to ensure the discrepancy sums to zero over a population of particles 
(in contrast to the adjusted target scheme), but the individual discrepancies can in theory be much larger than the previously
described schemes (although increasingly large
discrepancies represent more unlikely outcomes of the sampling). The discretized nature of the \cite{Sormani2017} scheme means that it
incurs a penalty to computational cost when sampling and increasing memory requirements when the IMF is sampled at finer
granularity (i.e. smaller mass bins), a problem not faced by other schemes in this section which allow for a continuum
of stellar masses.

\subsection{Monte Carlo tests of IMF sampling methods} \label{Monte Carlo}
\begin{figure*}
\centering
\includegraphics{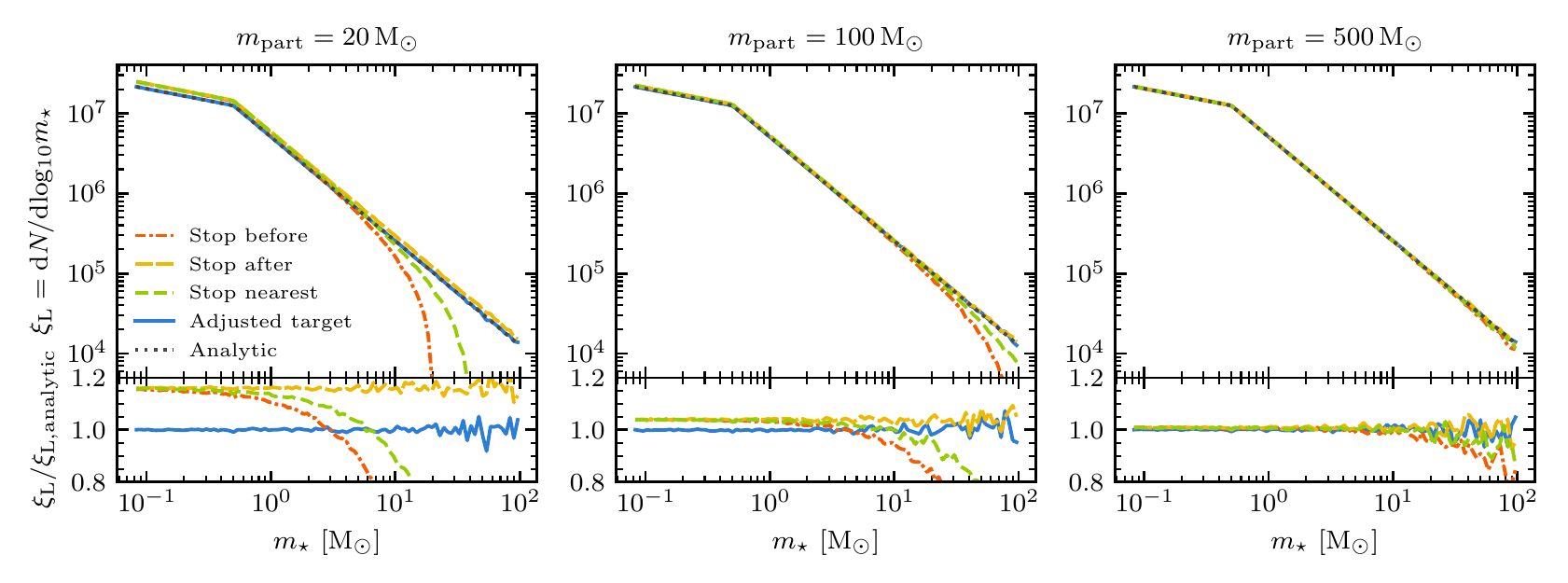}
\caption{The distribution of individual stellar masses when a mass budget of $10^7\,\Msun$ 
is divided into particles of mass $\mpart$
which are then populated using one of the four sampling methods described in the main text. The top panels show the the absolute
distribution while the bottom panels show the distributions relative to the analytic expectation. Both the stop before and
stop nearest methods bias the low-mass end of the IMF up and suppress the high-mass end. The stop after method preserves the
shape of the input IMF but results in a larger normalization across the whole mass range. These biases become less
pronounced for larger $\mpart$. The adjusted target scheme reproduces the input IMF for all $\mpart$.\vspace{-4ex}}
\label{fig_sample_hist}
\end{figure*}
\begin{figure*}
\centering
\includegraphics{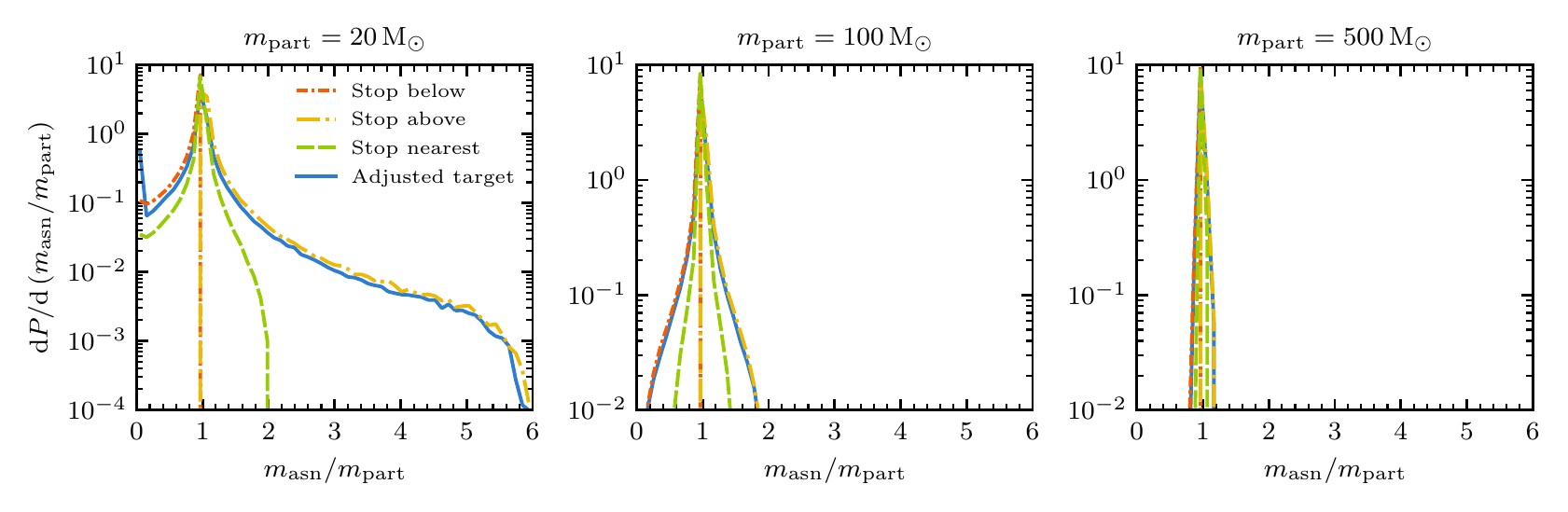}
\caption{PDFs of $m_\mathrm{asn}/\mpart$ for individual particles populated with stars using the four sampling methods. All
schemes produce a peak at unity i.e. where the assigned mass is completely consistent with the particle mass. Stop before
and stop have tails above and below unity, as expected, indicating particles where the assigned mass has under- or overshot,
respectively. The stop nearest scheme provides a tighter PDF. The adjusted target scheme essentially traces the wings of
the stop before and stop after schemes. The relative scatter about unity decreases as $\mpart$ is increased.\vspace{-4ex}}
\label{fig_mass_dist}
\end{figure*}
We now present some idealized Monte Carlo tests of the four main IMF sampling schemes described in the previous section
(stop before, stop after, stop nearest and adjusted target)
to demonstrate their properties when used to populate star particles. Throughout this work we will use a \cite{Kroupa2001}
IMF with minimum and maximum stellar masses of $m_\mathrm{min}=0.08\,\Msun$ and $m_\mathrm{max}=100\,\Msun$, respectively. Our findings will hold in a 
qualitative sense for any reasonable input IMF. We will draw samples from the IMF for a total mass 
budget of $10^7\,\Msun$. We have confirmed that this is a sufficiently large amount of mass that it can be populated
almost perfectly with stellar masses drawn from our input IMF (no matter what method is used) if no restrictions are applied.
However, we will instead subdivide this mass reservoir into particles of mass $\mpart$ and use the four IMF sampling
methods to populate them. This is representative of how star particles would be populated in a galaxy formation simulation.
We will use various values of $\mpart$ to examine the resolution dependence of the four schemes.

In Fig.~\ref{fig_sample_hist} we compare the total distribution of individual stellar masses across the various star particles to
the analytic expectation from the input IMF. Specifically, we show the number of stars per logarithmic mass bin, 
$\xi_\mathrm{L}=\mathrm{d}N/\mathrm{dlog_{10}}m_\star$. In the top panels we show $\xi_\mathrm{L}$ directly while in the bottom panels
we show its ratio to the analytic expectation from the input IMF. The general trends are as follows. 
The stop before method results in a suppression of the 
high-mass end of the IMF while boosting the low-mass end, as expected. The maximum stellar mass that 
can be accepted with the stop
before method is equal to the star particle mass, since any higher mass will carry it over the total. Even then, this is
only true when it is the first mass drawn for the particle. In general, because a more massive star is more likely to carry
the total assigned mass over the star particle mass than a less massive star, low mass draws are more likely to be accepted.
When $\mpart=20\,\Msun$, the shape of the output IMF is very distorted. There is a $\sim15\%$ enhancement 
of $\xi_\mathrm{L}$ for stars less than $\sim0.5\,\Msun$, followed by a sharp drop at higher masses with a net suppression
above $3\,\Msun$. The biasing effect is less severe as $\mpart$ is increased because the relative importance
of the final drawn stellar mass is reduced. However, there is still a strong suppression of the high-mass end of the IMF
when $\mpart=100\,\Msun$ and it is still noticeable when $\mpart=500\,\Msun$.

The stop after method always perfectly reproduces the shape of the input IMF (modulo some noise at the high-mass end) because
no draw is ever discarded. However, this always results in an overshoot of the particle mass meaning that the absolute number
of stars is biased upwards. When $\mpart=20\,\Msun$, this manifests as an enhancement of $\xi_\mathrm{L}$ by
an average of $16\%$ across the whole mass range. Again, this bias reduces as the particle mass increases, with an
enhancement of $3.9\%$ and $0.8\%$ when $\mpart$ is increased to $100\,\Msun$ and $500\,\Msun$, respectively.

The stop nearest method
suffers from the same issues as the stop before method, albeit to a lesser extent. In this method, a draw will be rejected if
the drawn mass is more than twice as large as the gap between the current total assigned mass and the particle mass.
It therefore also
enhances the low-mass end of the IMF while suppressing the high-mass end, although not as much as the stop
before scheme. Again, this bias is most pronounced at low particle masses and is 
only marginally apparent (but still present) when $\mpart=500\,\Msun$. Finally, Fig.~\ref{fig_sample_hist} shows
that the population across all star particles is highly consistent with the input IMF 
when the adjusted target scheme is used,
independent of star particle mass. It reproduces the shape and normalization almost perfectly, with only slight variation
at the high-mass end due to noise (since those stars are rare). The amplitude of the noise is the same as when the
total $10^7\,\Msun$ is sampled monolithically (not shown).

\begin{figure}
\centering
\includegraphics{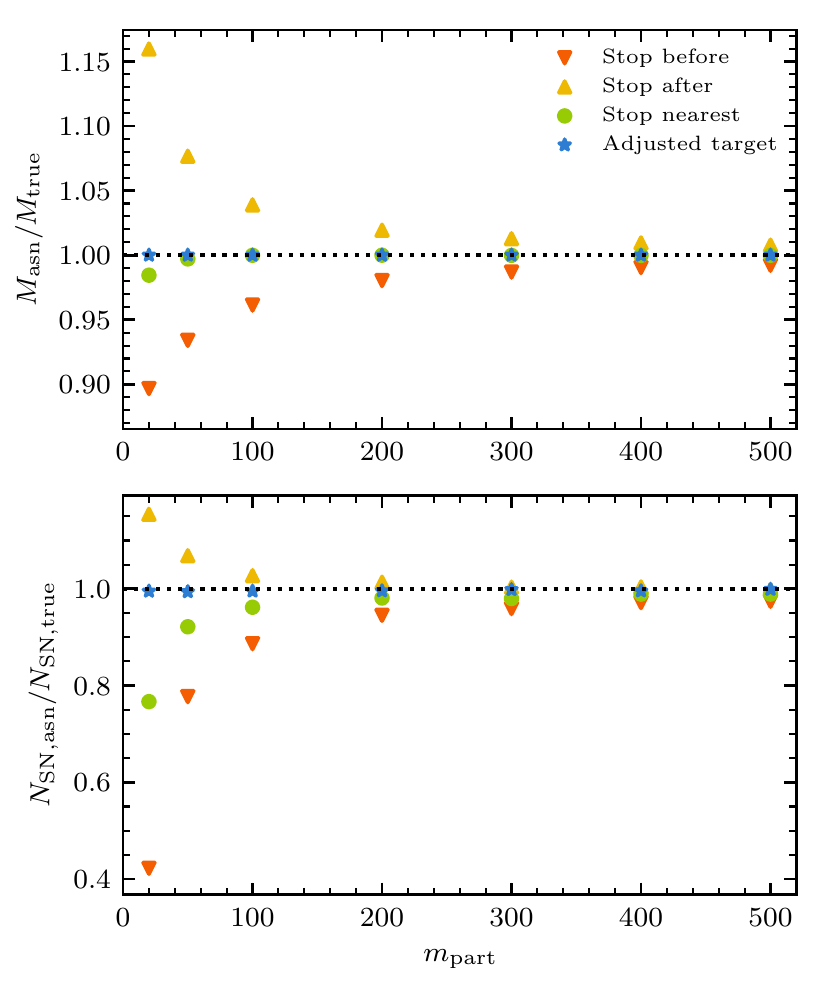}
\vspace{-6ex}
\caption{Top: Ratio between the total assigned mass and total star particle mass across the entire population of particles
for various particle masses, $\mpart$, and the four different sampling methods. Bottom: the total number of SN progenitors
in the populations relative to that predicted by the input IMF, assuming stars in the mass range $8-35\,\Msun$ produce
core-collapse SNe. Only the adjusted target scheme can produce a total mass in assigned stars that is consistent with the
star particle mass for $\mpart$ as low as $20\,\Msun$. It is also unique in producing the correct number of SNe progenitors
for all $\mpart$. The stop nearest approach has a deficit in the number of SN progenitors at low $\mpart$ but converges
to the correct value by $\mpart=500\,\Msun$.}
\label{fig_sample_bias} 
\end{figure}

As described in previous sections, IMF sampling schemes must balance preserving the input IMF with the two additional issues
of minimizing the inconsistency between the assigned and dynamical masses for each
individual star particle and conserving mass across the population of star particles. First, we shall examine the relative
mass inconsistency, $m_\mathrm{asn}/\mpart$, for the individual star particles populated in our Monte Carlo experiments.
Fig.~\ref{fig_mass_dist} shows PDFs for this ratio for our four trialled sampling methods with $\mpart=20\,\Msun$,
$100\,\Msun$ and $500\,\Msun$. All distributions peak at a value of 1 i.e. where the assigned mass is consistent with the
particle mass. The stop before has a
tail that extends to lower values of the ratio i.e. representing particles that have less assigned mass than their particle
mass. The reverse is true for stop after. These trends are an obvious result of the construction of the schemes. The stop
before scheme can assign no more mass than $\mpart$ and no less mass than 
$\mathrm{MAX}\left(\mpart-m_\mathrm{max},0\right)$. Equivalently, the stop after 
scheme can assign no less mass than 
$\mpart$ and no more mass than $\mpart + m_\mathrm{max}$. These limits can be seen
in the PDFs and result in a narrowing of the distribution as $\mpart$ is increased.

The stop nearest method results in the tightest distribution of $m_\mathrm{asn}/\mpart$ about unity of the four
schemes. It is
possible for the assigned mass to be below or above the particle mass. The minimum mass that can be assigned 
is $\mathrm{MAX}\left(\mpart-\frac{1}{2}m_\mathrm{max},0\right)$ while the maximum mass that can be assigned 
is $\mathrm{MIN}\left(2\mpart,\mpart+\frac{1}{2}m_\mathrm{max}\right)$. Again, the distribution tightens in a relative sense
as $\mpart$ increases because the impact of the last drawn stellar mass is reduced. The adjusted target scheme has
the largest spread in $m_\mathrm{asn}/\mpart$ of the four schemes. It too can assign more or less stellar mass than
the particle mass. Its minimum assignable mass is the same as the minimum for the stop before scheme and its maximum
assignable mass is the same as the maximum for the stop after scheme. It can be seen in Fig.~\ref{fig_mass_dist} that the
wings of the adjusted target PDFs trace those of the stop before and stop after PDFs. However, when $\mpart=20\,\Msun$,
adjusted target produces more particles with no assigned mass (relative to stop before and stop nearest) because there will
occasionally be a series of $m_\mathrm{asn}=0$ particles to compensate for a massive star being drawn previously (as
described in the previous section).

Fig.~\ref{fig_sample_bias} shows how the choice of sampling scheme affects the ratio between the total assigned mass for
the whole ensemble of particles, $M_\mathrm{asn}=\Sigma_j m_{\mathrm{asn},j}$, and the true mass budget,
$M_\mathrm{true}=10^7\,\Msun$, for various values of $\mpart$. As expected, the stop before scheme produces a deficit of 
assigned stellar masses
(because it always throws away the last draw) while stop after correspondingly produces an excess (because it always
keeps the last draw). This is most apparent at low values of $\mpart$, but a small bias is still evident 
at $\mpart=500\,\Msun$. Stop nearest has a marginal mass deficit at the lowest $\mpart$ but recovers the true mass by
$\mpart=100\,\Msun$. The adjusted target always recovers the true mass budget even with $\mpart=20\,\Msun$. Of particular relevance
to galaxy formation simulations is whether the total stellar feedback budget is recovered. Because the majority of the
budget is generated by massive stars, the recovery of the amount of feedback per unit stellar mass from the input IMF is
sensitive to how closely a sampling scheme reproduces the high-mass end of the IMF. In Fig.~\ref{fig_sample_bias} we also
show how the total number of SN progenitors generated by our Monte Carlo experiments compares to the expected number. We
assume that stellar masses in the range $8-35\,\Msun$ will produce core collapse SNe (we ignore Type Ia SNe). Again, as
expected, the stop before and stop after schemes produce too few or too many SN progenitors, respectively,
relative to the analytic expectation. The stop nearest scheme also produces too few SN progenitors with deficits of $23.8\%$,
$7.8\%$ and $3.8\%$ for $\mpart=20\,\Msun$, $50\,\Msun$ and $100\,\Msun$, respectively. It converges onto the correct number
of progenitors within $1\%$ for $500\,\Msun$. By contrast, the adjusted target scheme gives the correct number of SN
progenitors within $0.5\%$ or better across all the values of $\mpart$ we test. Note that if the maximum core collapse SNe
progenitor mass is increased, the biasing of the feedback budget will be worse. Likewise, feedback channels that depend 
strongly on the most massive stars (e.g. ionizing radiation) will show similar biases.

An examination of Fig.~\ref{fig_sample_hist} and \ref{fig_sample_bias} may suggest a simpler version of our adjusted target scheme.
Since the stop after scheme correctly recovers the shape of the IMF but has a constant offset across the entire mass range,
is it possible to recover the normalisation by choosing a single target mass for all particles 
(lower than the particle dynamical mass) in advance to compensate for the overshoot, rather than correcting from particle to
particle? We would set 
$m_\mathrm{target}= \left\langle \xi_\mathrm{L,\,analytic}/\xi_\mathrm{L}\right\rangle m_\mathrm{part}$,
where we have averaged the expected and output mass distributions from the 
stop after Monte Carlo experiment (Fig.~\ref{fig_sample_hist})
across the stellar mass range. 
This does indeed correctly compensate for the overshoot and restore the correct IMF normalisation.
However, it has several drawbacks. Firstly, while the normalisation is correct across many particles, over several consecutive
particles the discrepancy between the dynamical mass and the assigned mass is more significant than our adjusted target scheme. This is because this scheme produces a mass discrepancy resulting from uncorrelated over- and undershoots, meaning that
that the ratio between the total dynamical and assigned masses is a random walk about unity.\footnote{Note that the decision not to correlate over- and undershoots from one particle to another does not avoid any perceived `action at a distance'. This effect is simply an implicit
part of the renormalisation procedure, rather than being more explicit in the case of the adjusted target scheme. Regardless, as
previously mentioned, the choice to adopt an IMF a priori already implies an acceptance of some degree of `action at a distance'.}  By contrast, by design the adjusted target scheme always
pushes this ratio towards unity, resulting in smaller amplitude deviations
(e.g. by a factor of $\sim$3 for trials of 10 consecutively populated $20\,\Msun$ particles). This quality may be important
for regions with low SFRs in order to avoid significant under- or overestimations of the feedback budget.

Secondly, it is clear from Fig.~\ref{fig_sample_hist} and \ref{fig_sample_bias} that the magnitude of the correction factor
($\left\langle \xi_\mathrm{L,\,analytic}/\xi_\mathrm{L}\right\rangle$) depends on particle mass. It also depends on the IMF shape,
as well as the lower and upper bounds. 
Thus, a new correction factor must be obtained whenever these parameters are varied, either by performing another Monte Carlo experiment or by obtaining some functional form of this dependence. 
By contrast, our adjusted
target scheme is independent of both the particle mass and the details of the IMF. Thus, it adaptively copes with
variable IMFs with no modification (so long as the IMF does not change significantly between consecutive particles, instead varying smoothly with
time as, for example, local metallicity gradually changes). The simpler scheme would require additional calculation of the
correction factor from functional forms or lookup tables on-the-fly to treat an IMF that varied during a simulation. This is also likely to
increase the level of noise mentioned in the previous paragraph as the scheme relies on compensating over a larger ensemble of
particles. Additionally, even if a functional form for the dependence of the correction factor on star particle mass is obtained,
the simplified scheme requires the star particle mass to be uniform throughout the simulation. This is not the case for all
implementations of star formation. Many Lagrangian codes convert entire gas particles
to star particles, meaning that a scatter in the gas particle mass results in a scatter in star particle mass. While gas particle
masses may be uniform in the initial conditions, mass return from feedback processes or particle splitting/merging to achieve
variable
resolution will impart a mass scatter. 
As another example, in the pseudo-Lagrangian \textsc{Arepo} code (more details can be found in the next section),
(de-)refinement operations keep the gas cell mass to within a factor of 2 of a predetermined target. Converting cells into
star particles thus generates a scatter in the particle mass.
Our adjusted target scheme adaptively compensates for varying
particle mass.

To summarise, the stop before, stop after and stop nearest sampling methods all fail to reproduce the input IMF except at
large star particle masses. This has direct consequences for the stellar feedback budget, resulting in substantial under- or
overestimation of the number of SN progenitors when $\mpart$ is less than $\sim100\,\Msun$. By contrast, the adjusted target
scheme does not bias the input IMF in any way, produces a total assigned mass that is consistent with the total star particle mass and recovers the correct feedback budget. The sole advantage of the stop nearest method is that it is the best
scheme for minimizing the discrepancy between the assigned and dynamical mass of an individual particle. However, this
discrepancy is not particularly important as galaxy formation simulations cannot typically resolve exact N-body dynamics and
the much larger inconsistency with the input IMF and the total stellar mass budget is a far worse penalty.
Above $\mpart=500\,\Msun$ the schemes largely converge, so
in this mass range the stop nearest approach may be adopted simply because it can be implemented locally to each task in a parallel computation. However, implementing our improved scheme is not substantially more complex and carries negligible
additional computational penalty. Our scheme can be integrated into any pre-existing star formation scheme that uses star
particles with a minimum of effort because it does not require additional modification of the mass of star particles or the
transfer of mass between star particles.

\vspace{-4ex}
\section{Explicit IMF sampling vs. IMF averaging in simulations} \label{imfex vs imfav}
In this section we will study the impact on galaxy evolution simulations of using explicit 
IMF sampling to determine the stellar feedback budgets of star
particles as opposed to the more common IMF averaged approach. We will present simulations of an idealized isolated
dwarf galaxy with a baryonic mass resolution of $20\,\Msun$. We include stellar feedback in the form of SNe, photoionization and
photoheating in \ion{H}{ii} regions and photoelectric heating of dust grains. Detailed descriptions of these models can be found in 
\cite{Smith2020} \citepalias[hereafter][]{Smith2020}, 
but we will summarise the salient details in the following section.
For convenience, we will denote simulations
with SNe, ionizing radiation and photoelectric heating by the abbreviations \simSN{}, \simPI{} and \simPE, respectively.
Simulations with all feedback channels switched on are denoted as \simSNPIPE{} while \simNoFB{} is adopted when no feedback is used.
We will refer to simulations
that explicitly sample the IMF by the abbreviation \mbox{`\textit{IMFsam}'} and simulations that use IMF averaged stellar feedback
as \mbox{`\textit{IMFav}'}. The \textit{IMFsam} simulations were originally presented in \mbox{\citetalias{Smith2020}}.

\vspace{-4ex}
\subsection{Numerical methods} \label{Sim methods}
\vspace{-1ex}
We use the moving-mesh code \textsc{Arepo} 
\citep{Springel2010,Pakmor2016} along with our own sub-grid models for star formation and stellar feedback 
\citepalias[which are described in more detail in][]{Smith2020}.
We use the \textsc{Grackle} chemistry and cooling library\footnote{https://grackle.readthedocs.io} \citep{Smith2017} in
its primordial six-species non-equilibrium mode, along with tabulated metal cooling, ionization and heating from
a meta-galactic UV background \citep{Haardt2012} and the self-shielding prescription of \cite{Rahmati2013}.
Gas cells can have a non-zero SFR when their Jeans mass drops below 8 times the cell mass. The SFR is then given
by a simple Schmidt law, $\dot{\rho}_\star = \epsilon_\mathrm{SF}\rho/t_\mathrm{ff}$, 
where $\rho$ is the gas density,
$t_\mathrm{ff}=\sqrt{3 \pi / 32 G \rho}$ is the local free-fall time and we adopt a fixed efficiency of
$\epsilon_\mathrm{SF}=0.02$, motivated by observed efficiencies in dense
gas \citep[see e.g.][and references therein]{Krumholz2007}. The SFR is then stochastically sampled to convert gas cells
into collisionless star particles.

We include photoelectric heating of dust grains by a spatially varying far-UV (FUV) field generated by star particles. 
The FUV energy density at each location in the domain is calculated using the gravity tree to sum the fluxes from sources,
using a local approximation for attenuation by dust which is valid in dust-poor systems. We approximate the dust-to-gas ratio
as a function of metallicity using the broken power-law of \cite{RemyRuyer2014}. \ion{H}{ii} regions around ionizing
sources are included using a novel anisotropic overlapping Str{\"o}mgren type approximation \citepalias[first presented in][]{Smith2020}. 
The balance between the
ionizing photon luminosity and the recombination rate is calculated in independent angular pixels around sources to
determine the extent of an \ion{H}{ii}. This
helps mitigate the mass-biasing error encountered by previous methods. The algorithm accounts for \ion{H}{ii} regions
from multiple sources overlapping. If a cell is tagged as belonging to an \ion{H}{ii} region, it is immediately heated
to $10^4$~K and is forbidden from cooling below that temperature while it remains tagged. SN feedback is included with the scheme
first presented in \cite{Smith2018}. This injects mass, metals, energy and momentum into the gas cell containing a star
particle and its immediate neighbours (those that share a face with the host cell). The scheme ensures an isotropic
injection of feedback quantities which is non-trivial in a Lagrangian code. We use the SN scheme in its mechanical feedback
mode which compensates for missing momentum when the Sedov-Taylor phase of a SN remnant is unresolved. In practice, at the
resolution we adopt in this work the majority of SNe are well resolved and we find that we achieve similar results when
a simple thermal dump of SN energy is used. In this work, we do not include stellar winds.

When the explicit IMF sampling scheme is used (\mbox{\textit{IMFsam}}), 
star particles are populated with an inventory of stellar masses 
at the moment of creation using the adjusted target scheme described in Section~\ref{pop IMF}. Rejection sampling is used
to draw stellar masses from the IMF. We use a single value of the
offset mass for the whole galaxy at any one time, rather than concurrent versions for different regions of the galaxy
(see the discussion in Section~\ref{pop IMF methods}).\footnote{Our simulations are parallelised using MPI, so an effort must be
made to synchronize the sampling between tasks (i.e. there must be a well defined order of star particle sampling). In
practice, we find the simplest and most computationally efficient way to achieve this is to perform all sampling on a single
MPI rank with the resulting inventories distributed back to the ranks on which the star particles reside.} We do not explicitly
record the masses of stars less than $5\,\Msun$ since they would contribute negligible feedback (as we do not include winds from
AGB stars),
although they are taken into account for the purposes of determining the total stellar mass assigned to the star
particle. For stars more massive than $5\,\Msun$, we then use lookup tables as a function of mass in order to determine
quantities relevant to feedback. We make the simplifying assumption in this work that all stars have a metallicity
of $0.1\,\Zsun$, the initial metallicity of our initial conditions. None of the quantities used have a strong
dependence on metallicity relative to the deviation from the initial metallicity that we see in this work, so this
approximation is reasonable.
We obtain the lifetime of the star from the PARSEC grid of stellar tracks 
\citep{Bressan2012}. The FUV and ionizing photon luminosities of the star are derived from the OSTAR2002 
grid of stellar models \cite{Lanz2003} as compiled by \cite{Emerick2019}, making the approximation that they are
fixed at their main sequence values throughout the life of the star. The net luminosity of a star particle is the sum of
all extant stars assigned to it. When a star reaches the end of its life it ceases to contribute radiation. 
Star particles with extant massive stars have their time-steps limited to 0.1~Myr, although in practice their time-step is
usually much smaller as constrained by other criteria. If a star in the
mass range $8-35\,\Msun$ reaches the end of its life, a SN is triggered, resolved from the host star particle. The ejecta mass and metallicity
depends on the progenitor mass, based on \cite{Chieffi2004}. We use a constant SN energy of $10^{51}$\,ergs for all SNe.
When SN feedback is switched off, we still return ejecta when a SN progenitor reaches the end of its life but do not
inject the energy.

For simulations using IMF averaged feedback quantities (\mbox{\textit{IMFav}}) the sampling procedure is not performed.
Instead, lookup tables of FUV and ionizing photon luminosities, and SN rates per unit stellar mass 
as a function of population age are used.
For consistency, these tables are derived using the same relationships between 
individual stellar mass and feedback quantities that are used in the \textit{IMFsam} schemes. We begin by populating
a $10^8\,\Msun$ SSP with stellar masses drawn from the IMF, the high mass budget guaranteeing that the IMF is very well sampled. 
Lifetimes and luminosities are obtained in a similar manner to the on-the-fly approach used in the \textit{IMFsam} schemes.
We then integrate through the lifetime of the SSP, recording how the net FUV and ionizing photon luminosities decrease as
stars reach the end of their lives. We also obtain SN rates by recording when stars in the $8-35\,\Msun$ mass range die.
We then normalize these luminosities and rates by the initial mass of the SSP. 
When a simulation is performed, star particles
are assigned their luminosities using these lookup tables based on their age. The expected
number of SNe that will be produced by a star particle in a given time-step, $\Delta t$, is
\begin{equation} 
\bar{N}_\mathrm{SN}=\dot{n}_\mathrm{SN}\left(t_\mathrm{part}\right)m_{\mathrm{part},0}\Delta t, 
\end{equation}
where $\dot{n}_\mathrm{SN}\left(t_\mathrm{part}\right)$ is the SN rate per unit mass obtained from the lookup tables as function of star particle age and $m_{\mathrm{part},0}$ is the initial star particle mass at the point of creation,
before mass loss due to feedback. The SN rate is normalized by $m_{\mathrm{part},0}$ rather than
$\mpart$ in order to recover the normalization used when creating the rate tables from the SSP. In other words, SNe
are independent events representing an implicit sampling of the IMF and 
so the chances of a SN being generated from a star particle should not be reduced
because a previous SN occurred and reduced the mass of the particle. Likewise, luminosities obtained from the lookup
tables are normalized by the initial particle mass, not the current mass. In constructing the IMF averaged luminosities
we have already explicitly linked the radiation output to the initial mass of the SSP and it is therefore
independent of the SN events that have occurred in the star particle and the ensuing mass loss.
The number of SNe that occur in a given time-step is then determined by drawing
from a Poisson distribution with a mean $\lambda=\bar{N}_\mathrm{SN}$. In Section~\ref{Discussion} we will discuss in more
detail why Poisson sampling is the appropriate procedure. If a SN occurs, we map the age of the star particle back onto
the mass of the progenitor to obtain the ejecta mass and metallicity. Like the \textit{IMFsam} schemes, star particles have 
a maximum time-step of 0.1~Myr enforced. This ensures the changes of luminosity are time-resolved, but also guarantees that
$\bar{N}_\mathrm{SN} \ll 1$. This is necessary to ensure that SNe are individually time-resolved, which is important to accurately
capture clustering effects.

It is possible that a star particle may not have sufficient mass to return as ejecta for a triggered SN event. For the \textit{IMFsam}
scheme this is a result of the potential discrepancy between the assigned and dynamical mass of the star particle. For the
\textit{IMFav} scheme this is because the stochastic sampling of the rates means that it is possible,
albeit unlikely, for a single star particle to produce more SNe than its mass should allow. If there is insufficient mass
in a star particle to return as ejecta, we instead return as much as possible (i.e. the mass of the particle) and remove the
particle. As reported in \citetalias{Smith2020}, this is a relatively rare occurrence and results in a overall reduction
of the total ejecta across all SNe by 5.1\% for a $20\,\Msun$ star particle. Only $2\%$ of SNe have
their ejecta mass reduced by more than 30\% and none have their ejecta reduced by more than 36\%. 
This may be of concern if a detailed
study of chemical enrichment is of interest (ignoring uncertainties in yields) but for this work we find this level
of deviation from the tabulated yields acceptable. It is also possible that in the event of an entire star particle being
removed due to a lack of ejecta mass another massive star hosted in the particle is prematurely 
deleted before it has reached the end of its life. This occurs in less than 0.1\% of the SN events in our simulations.
\vspace{-4ex}
\subsection{Simulation details}
For our idealized dwarf galaxy we use the `fiducial' initial conditions from \citetalias{Smith2020},
generated using the \textsc{MakeNewDisk} code \citep{Springel2005a}. The system has a
virial mass of $10^{10}\,\Msun$. There is a $6.825\times10^7\,\Msun$ gas disc with an exponential radial density profile with a scale length of 1.1~kpc. The vertical structure is set to achieve hydrostatic equilibrium at an 
initial temperature of $10^4\,\mathrm{K}$.
The gas is initialized with a metallicity of $0.1\,\Zsun$. We do not include a circumgalactic medium (CGM). There is also
an initial stellar disc with a mass of $9.75\times10^{6}\,\Msun$ and the same radial density profile as the gas disc.
The vertical structure is Gaussian with a scale height of 0.7~kpc. Star particles present in the initial conditions do
not contribute stellar feedback. The rest of the mass of the system is in the form of a live, spherically symmetric 
dark matter halo. This follows a \cite{Hernquist1990} density profile chosen to provide a close match to a
\cite{Navarro1997} profile with a concentration parameter, $c$, of 15 and a spin parameter, $\lambda$, of 0.04.
Gas cells and star particles have a mass of $20\,\Msun$ (derefinement and refinement operations keep the gas cells within
a factor of 2 of this target mass), while dark matter particles have a mass of $1640\,\Msun$. Gravitational softening lengths
are fixed at 1.75~pc and 20~pc for star and dark matter particles, respectively. Gas cells use adaptive softening lengths
down to a minimum of 1.75~pc. The initial conditions are relaxed for 100~Myr with cooling but without star formation while
initial turbulence is driven with a pseudo-SN feedback scheme \citepalias[described in detail][]{Smith2020}. This is to avoid a rapid
vertical collapse of the disc when the simulations are started.
\vspace{-4ex}
\subsection{Results} \label{results}

\begin{figure}
\centering{}
\includegraphics{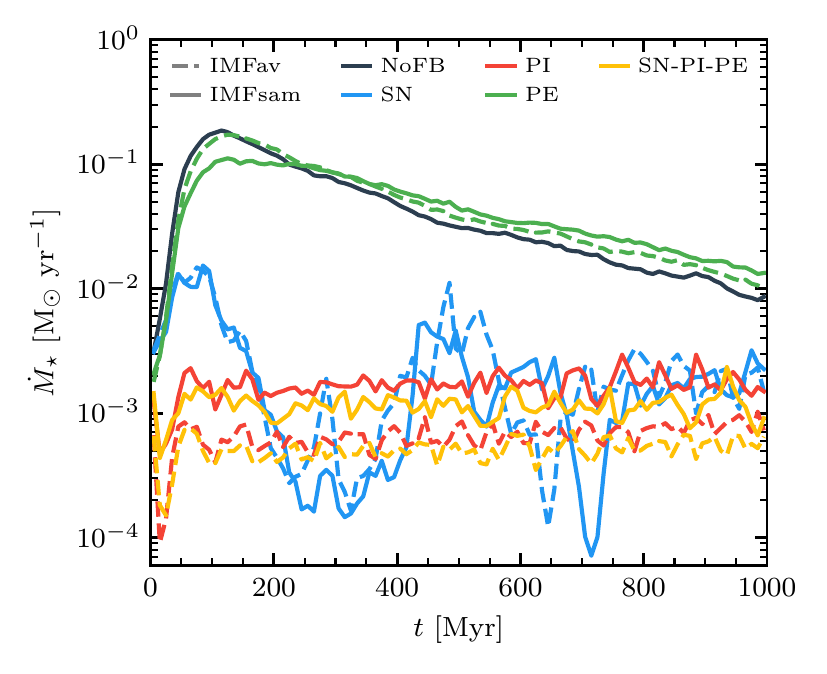}
\vspace{-6ex}
\caption{SFRs for our simulations, averaged over 10~Myr. Simulations without feedback or with photoelectric heating alone
result in a high SFR limited only by the available gas supply. SN feedback regulates SF in a bursty manner with
IMF averaging (\textit{IMFav}) and explicit IMF sampling (\textit{IMFsam}) giving essentially identical results.
Photoionization feedback is also able to regulate star formation. However, it is more efficient by a factor of 2.4 in 
\textit{IMFav} compared to \textit{IMFsam}.\vspace{-4ex}}
\label{fig_sfr} 
\end{figure}

Fig.~\ref{fig_sfr} shows the SFR for the simulations, averaged over 10~Myr. Dashed lines show results from 
simulations with IMF averaged feedback (\textit{IMFav}) while solid lines show simulations with explicit IMF sampling
(\textit{IMFsam}). This convention is used throughout this work. Without feedback, the SFR rises rapidly and is only limited
by the supply of gas (we only perform an \textit{IMFsam} \simNoFB{} simulation). When feedback is included, the results
are qualitatively the same between \textit{IMFav} and \textit{IMFsam} simulations. A detailed discussion about how the various
feedback schemes regulate the SFR can be found in \citetalias{Smith2020}. As found in that paper, photoelectric heating by itself 
is inefficient,
providing little suppression of star formation. It is slightly more efficient in the \textit{IMFsam} simulation, reducing the
peak SFR reached, but this is an extremely marginal effect. SN feedback is able 
to suppress the SFR by almost two orders of magnitude
on average. It does so in a very bursty manner. The \textit{IMFav} and \textit{IMFsam} produce near identical results.
Differences between the two are well within the margin of numerical noise arising from stochastic sampling/triggering of
SNe. We confirm this in Appendix~\ref{stoch}, where we repeat the early stages of 
these simulations four additional times with different seeds
for the random number generators and with randomly perturbed initial conditions.

Photoionization feedback is also able to regulate SFRs (see \citetalias{Smith2020} for a detailed discussion of why this occurs). 
It produces a smoother SF history as it disrupts star forming
regions locally in a more gentle manner. When averaged over the last 500~Myr of the 
simulation (allowing the initial transient to settle), 
the \textit{IMFsam} simulation 
regulates the SFR to roughly the same degree as the SN feedback simulations. However, the \textit{IMFav} reduces the SFR
by a factor of 2.4 relative to the \textit{IMFsam} simulation. When all feedback channels are turned on (\simSNPIPE) with
explicit IMF sampling, SFRs are reduced by $\sim40\%$ relative to the SN or photoionization only simulations,
but this is small compared to the initial reduction from the no feedback case. When the IMF averaged values are used instead,
switching on all the feedback results in a reduction of 26\% relative to the \textit{IMFav} \simPI{} simulation but 73\% relative to the \textit{IMFav} \simSN{} simulation. Photoionization is therefore the dominant regulator of star formation
in the \textit{IMFav} simulations, in contrast to the \textit{IMFsam} case where we observe parity. The relative difference
between the strength of photoionization feedback between the \textit{IMFav} and \textit{IMFsam} is not a stochastic effect,
as can be seen in Appendix~\ref{stoch}.

\begin{figure}
\centering{}
\includegraphics{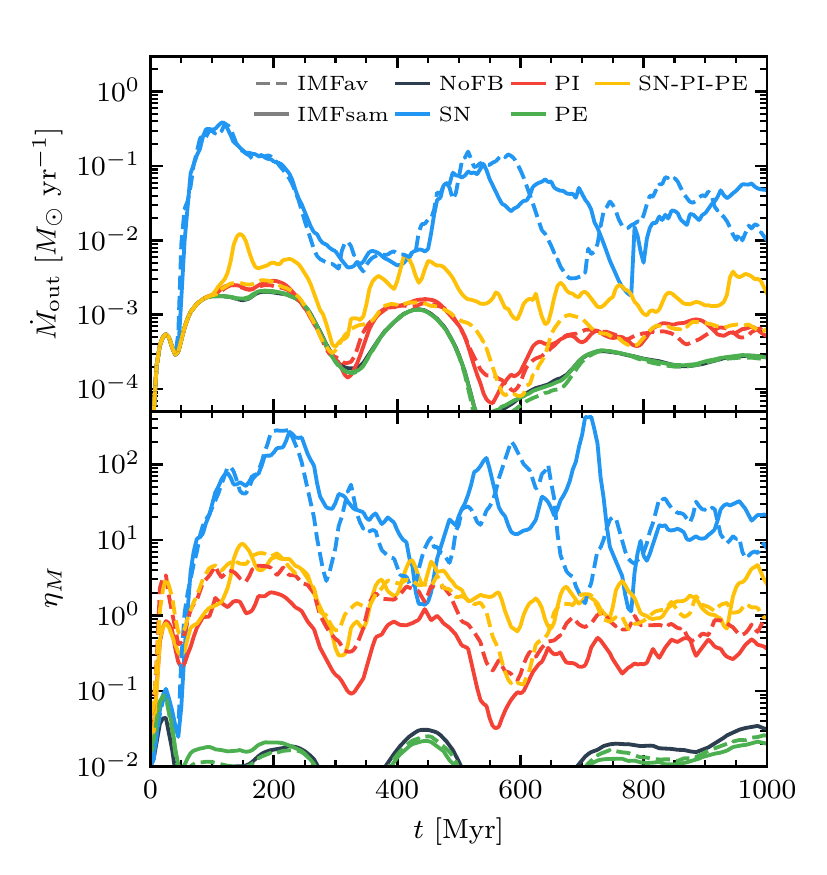}
\vspace{-6ex}
\caption{Outflow rates across thin slabs parallel to the disc at 1 kpc above and below the disc midplane. The top panel shows the
absolute rates while the bottom panel normalises by the SFR (as in Fig.~\ref{fig_sfr}) to produce mass loading
factors. In line with the findings of \protect\citetalias{Smith2020}, the addition of photoionization feedback
diminishes SN-driven winds in these simulations. This effect is more pronounced for the IMF averaged feedback case.
\vspace{-4ex}}
\label{fig_outflow} 
\end{figure}
\begin{figure}
\centering
\includegraphics{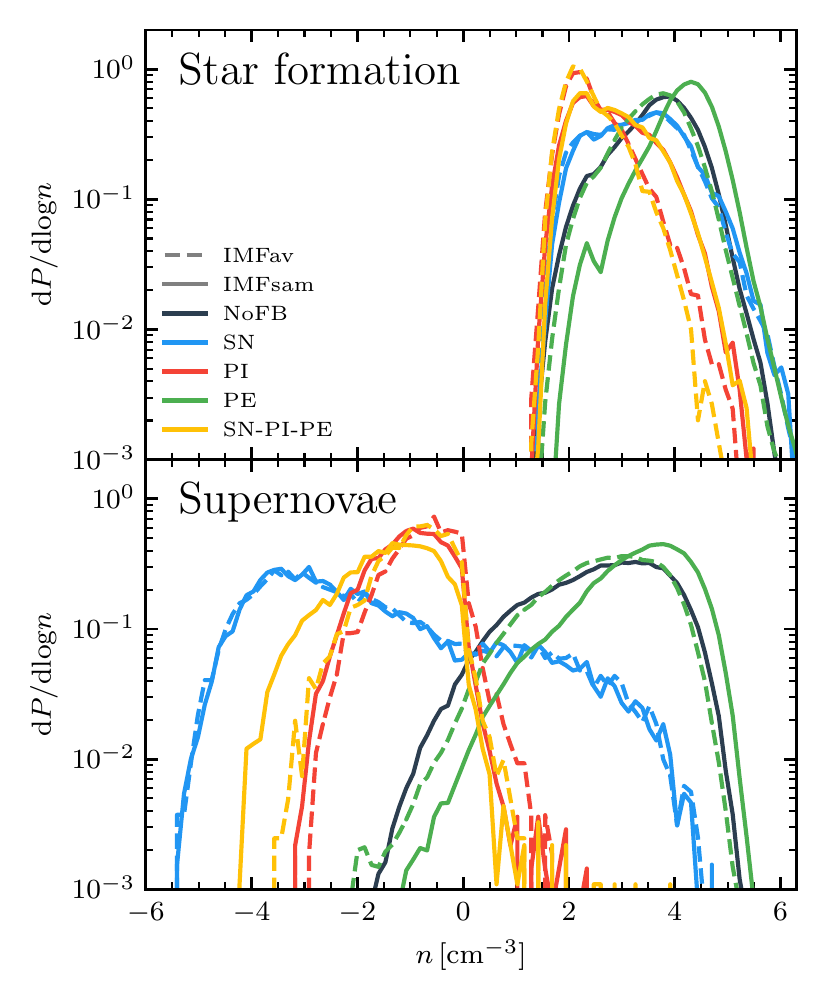}
\vspace{-6ex}
\caption{PDFs of the ambient density where star particles are created (top) and SNe explode (or would explode if SN
feedback was switched on) (bottom), ignoring the first 400 Myr. SNe broaden the distribution of birth densities. Ionizing
radiation shifts the peak to lower densities as it prevents the collapse of gas to high densities. This effect is more
pronounced in the \textit{IMFav} simulations. When SNe are the only source of feedback, SNe explode over a broad range of
densities as dense clouds are dispersed and superbubbles are created. \textit{IMFav} and \textit{IMFsam} give identical
results. When ionizing radiation is included, dense gas is dispersed prior to SNe occurring but the growth of superbubbles
is restricted, preventing the low density tail from extending as far. It extends slightly further in \textit{IMFsam}
\simSNPIPE{} because the effect of ionizing radiation is not as strong.\vspace{-4ex}}
\label{fig_sfsn_pdfs} 
\end{figure}
\begin{figure*}
\centering
\includegraphics{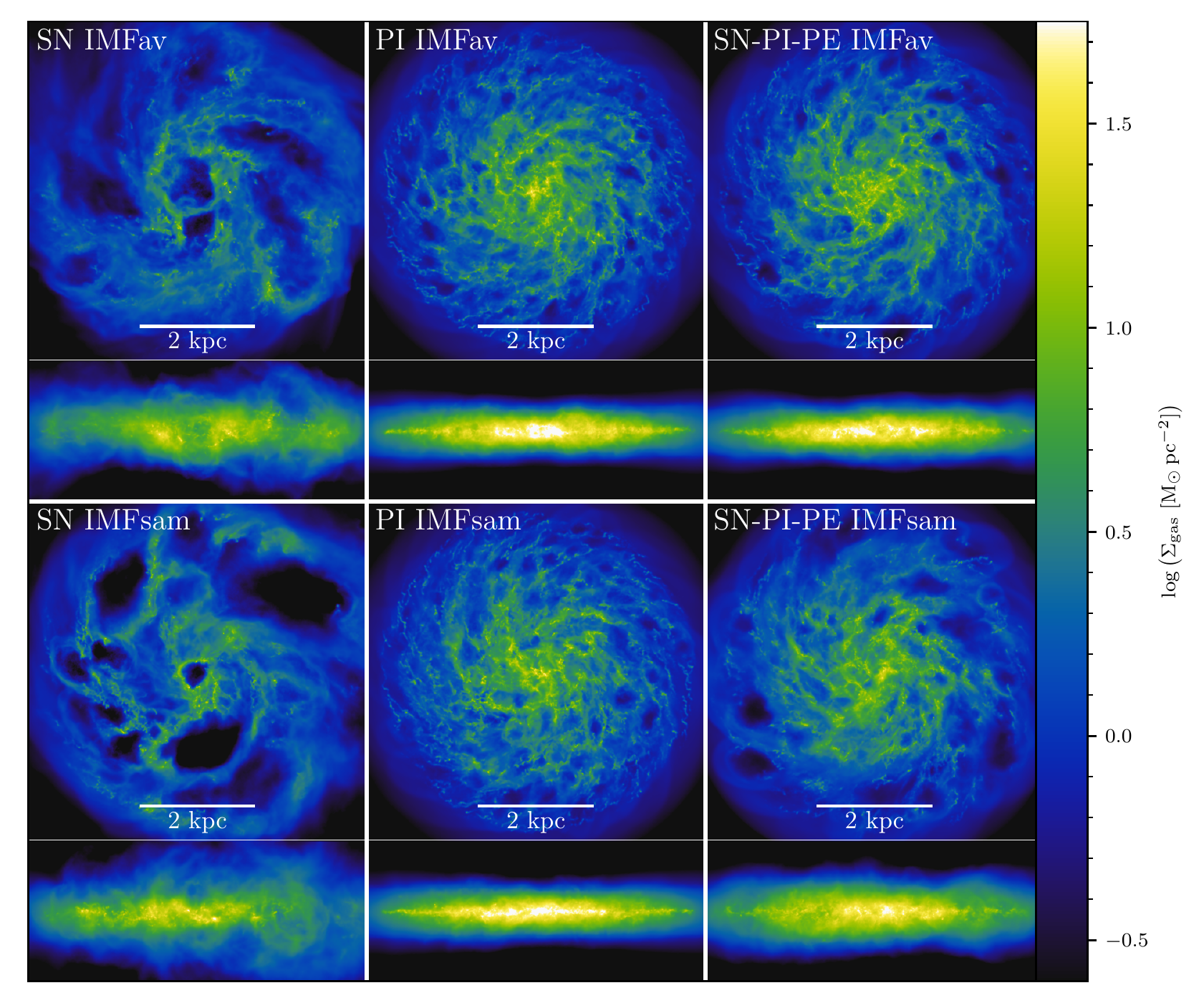}
\caption{Face-on and edge-on projections of the gas discs at 1~Gyr. \simNoFB{} and \simPE{} simulations are not shown.
SN feedback alone produces a thick and highly disordered disc, with large transient cavities caused by SN superbubbles.
When ionizing radiation is the only source of feedback the discs are more ordered and the large holes are not evident.
The \simSNPIPE{} \textit{IMFav} simulation is qualitatively the same as the \simPI{} \textit{IMFav} run, but when
explicit IMF sampling is used small, transient SN-driven cavities appear marginally more frequently.}
\vspace{-4ex}
\label{fig_proj} 
\end{figure*}
\begin{figure*}
\centering
\includegraphics{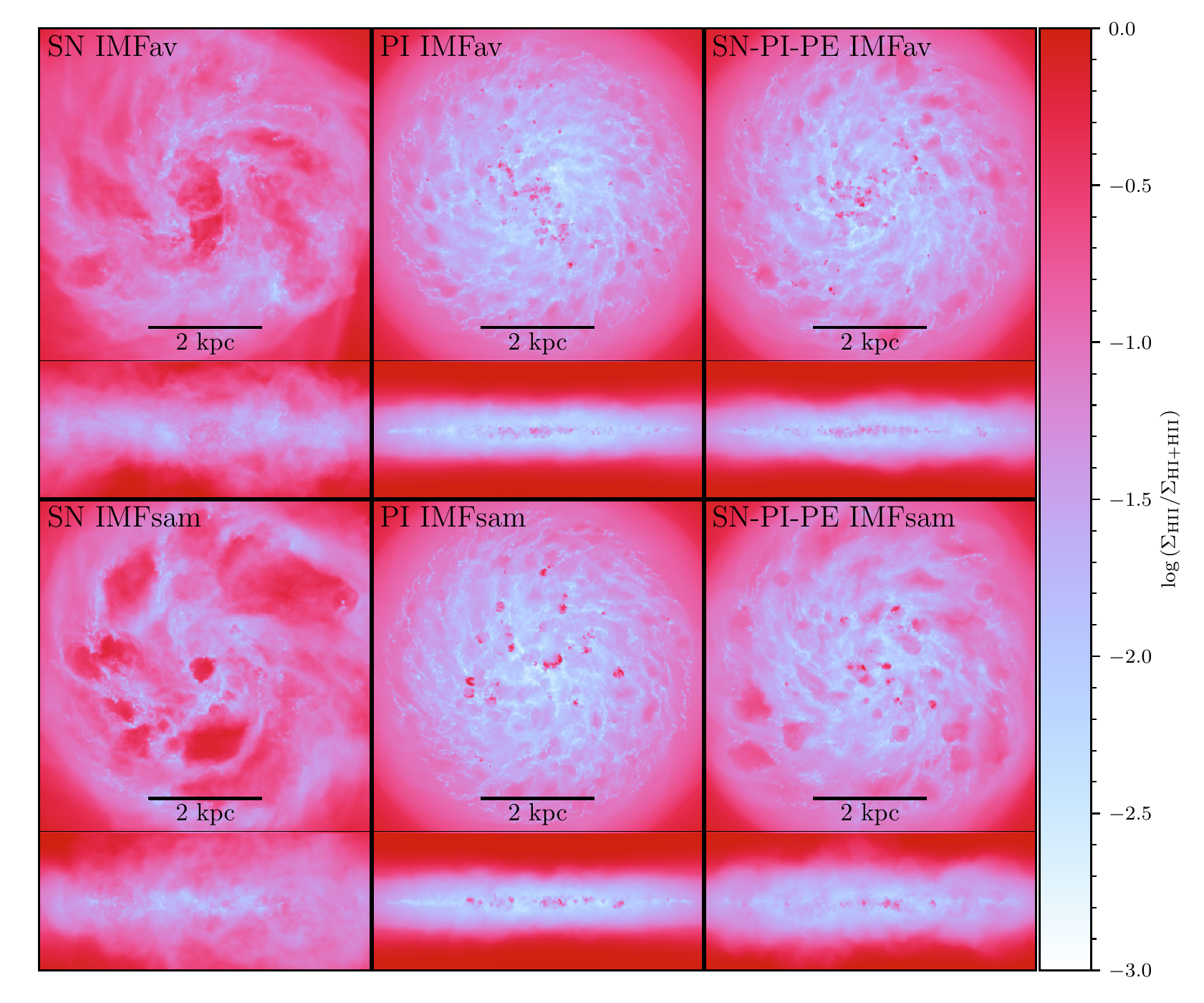}
\caption{The surface density of \ion{H}{ii} divided by the total hydrogen density at 1~Gyr, shown face-on and edge-on.
\simNoFB{} and \simPE{} simulations are not shown. Generally, the ionized regions trace the low density gas. Patches of
high ionization fraction can be seen embedded in dense gas in the simulations with photoionization feedback. These are
\ion{H}{ii} regions around ionizing sources. The \textit{IMFav} simulations produces many small \ion{H}{ii} regions
scattered throughout the disc compared to the \textit{IMFsam} approach which features rarer, but larger \ion{H}{ii} regions.}
\vspace{-4ex}
\label{fig_proj_hii} 
\end{figure*}

{
Fig.~\ref{fig_outflow} shows the mass outflow rate measured through parallel slabs 1~kpc above and below the disc midplane.
This is measured as:
\begin{equation}
\dot{M}_\mathrm{out} = \frac{\sum_{i}\left( m_i v_{\mathrm{out},i}\right)}{\Delta z},
\end{equation}
where the summation is carried out over all gas cells with a positive outflow velocity, $v_\mathrm{out}$, perpendicular to the
disc plane located within a slab of thickness $\Delta z = 100\,\mathrm{pc}$. Fig.~\ref{fig_outflow} shows both the absolute rates
and mass loading factors. The latter is obtained by normalising the absolute rates by the SFR averaged over the preceding 10~Myr
as shown in Fig.~\ref{fig_sfr}. The results for the \textit{IMFsam} simulations were presented and discussed at length in
\citetalias{Smith2020}. We refer the reader to that work for a detailed explanation of the impact of various combinations of
feedback channels on outflow rates, but give the salient details here. In the absence of efficient feedback (i.e. the \simNoFB{} or \simPE{}
simulations)
there is a small flow of gas outwards through 1~kpc, arising from the settling of the idealised initial conditions (note that
there is no CGM). 
The addition of photoionization feedback leads to a small enhancement of this outward flow due to 
additional thermal support and the momentum input from expanding \ion{H}{ii} regions. The impact on the apparent mass loading 
factor is large due to the significant reduction in the SFR. However, because the outflow is only a minor enhancement of
a flow that existed in the \simNoFB{} simulation, caution should be adopted before interpreting 
the mass loading factor as indicating the strength of a feedback
driven wind in this case. 

SN feedback alone leads to strong bursts of outflows with mass loading factors of 10-100 for both
\textit{IMFav} and \textit{IMFsam} simulations.
However, combining all the feedback channels (\simSNPIPE{}) leads to a suppression of outflow rates
by roughly an order of magnitude. As demonstrated in \citetalias{Smith2020}, this is due to a reduction in the clustering
of SNe in both space and time by the pre-SN photoionization feedback. 
This effect is significantly more pronounced for the \textit{IMFav} case, with essentially no enhancement of absolute outflow
rates relative to the simulations without SNe and a significant deficit relative to the equivalent \textit{IMFsam} run.
The reduction in absolute outflow rate is mainly a result of the lower overall SFR in these runs, leading to a correspondingly
lower SN budget to drive outflows. We might also expect a reduction in the loading factor due to even more significant reduction
in SN clustering, but it is comparable between \textit{IMFav} and \textit{IMFsam}. Again, this is because in the absence of
efficient SN feedback the mild outflow we see across 1~kpc is only weakly related to stellar feedback, so the variation of the mass
loading factor is dominated by the differing SFR. Note that the mass loadings for the \simSN{}, \simPI{} and \simSNPIPE{} simulations
(with both IMF schemes) are all consistent with the range provided by observations of outflows from dwarfs \citep[see e.g.][]{McQuinn2019}, although strong mass loadings from low mass galaxies are often required by theory \citepalias[again, we refer
the reader to the discussion in][]{Smith2020}.
}

Fig.~\ref{fig_sfsn_pdfs} shows PDFs for the ambient density where star particles are created and where SNe occur\footnote{
In simulations where SN feedback is switched off, we still record where SNe would occur either based on a SN progenitor
reaching the end of its life (in the \textit{IMFsam} simulations) or by sampling the SN rates (in the \textit{IMFav}
simulations). As noted in Section~\ref{Sim methods} we also return ejecta mass but do not add SN energy.
}. We exclude the first 400~Myr to ignore the initial
transient phase of the simulation. Our Jeans mass based star formation criteria means that the onset of star formation
occurs between $\sim20-100\,\mathrm{cm^{-3}}$. Simulations without feedback or with photoelectric heating alone peak at
around a density of $10^4\,\mathrm{cm^{-3}}$ as gas continues to collapse beyond the onset of star formation. The
difference between the \textit{IMFav} and \textit{IMFsam} photoelectric heating runs is 
simply a result of the different gas fractions in the disc due to
the offset in SFR peaks in the first 400~Myr, essentially producing an offset of the disc evolution in time.
When SN feedback is used alone, the PDF of star formation ambient densities
is broadened. This is because the feedback alters the distribution of dense gas by driving turbulence and disrupting
collapsing clouds. \textit{IMFav} and \textit{IMFsam} give identical PDFs. When ionizing radiation is included the peak of the
distribution is moved to lower densities and the maximum density reached is reduced. This is because the radiation is able
to halt the collapse of dense clouds and disrupt them earlier than the SN feedback (which is delayed by the lifetime of
its progenitors). It can be seen that this effect is significantly more pronounced in the \textit{IMFav} 
than the \textit{IMFsam} simulations, a key difference between the two methods.

In the absence of feedback or with photoelectric heating alone, the PDFs of the ambient density where SN progenitors die
reflects the PDFs of star particle birth densities with a broadening towards lower densities. This lower density tail arises
because of runaway gas consumption in star forming clouds, dropping the local density, as well as being caused by star
particles drifting out of their (now very compact) birth clouds. When SN feedback is used alone, the PDF of SN site densities
spans roughly ten orders of magnitude in density. The first SNe to occur in a star forming cloud explode in the dense gas of
star forming regions. They are able to disperse these dense clouds and successive SNe contribute to the creation of a
superbubble, with each subsequent SN occurring in lower density gas. This gives rise to the broad range of ambient densities.
Again, the simulation using IMF averaged SN rates and that explicitly sampling the IMF for SN progenitors produce essentially
identical PDFs. When ionizing radiation is included (either on its own or with the other feedback channels) 
almost no SNe occur in star forming gas. The radiation is able to clear dense gas prior to SNe occurring. The PDFs are similar
for the \textit{IMFav} and \textit{IMFsam} simulations. The low density tail does not extend as far as the 
SN only simulations. This is because the creation of superbubbles is inhibited by the ionizing radiation acting to reduce
the clustering of SNe, as shown in greater detail in \citetalias{Smith2020}. The \simSNPIPE{} \textit{IMFsam} PDF extends
to slightly lower densities than the \textit{IMFav} equivalent because the impact of the radiation on clustering is not
as strong in this case, allowing some SNe to occur in larger bubbles.

\begin{figure}
\centering
\includegraphics{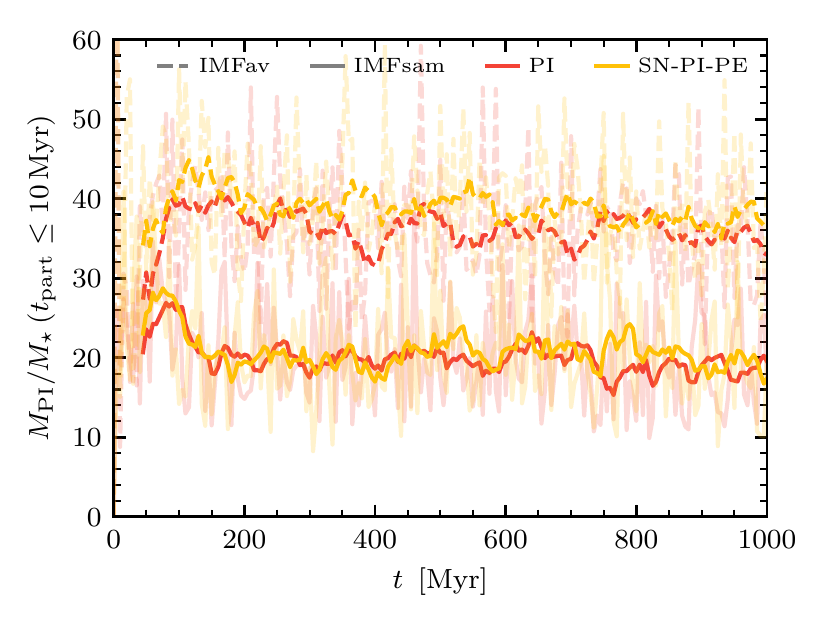}
\vspace{-6ex}
\caption{The mass of gas tagged as belonging to an \ion{H}{ii} region normalized by the mass in young stars. The
instantaneous value (shown in the paler colours) is very noisy so we also plot a 50~Myr moving average (the bold colours).
The \textit{IMFav} simulations produce almost twice as much photoionized gas per unit stellar mass as the \textit{IMFsam}
equivalents, because sources are typically embedded in lower density gas.}
\vspace{-4ex}
\label{fig_ionmassloading} 
\end{figure}

Fig.~\ref{fig_proj} shows face-on and edge-on projections of the gas disc after 1~Gyr. Simulations \simNoFB{} and \simPE{}
are not shown, but equivalent plots can be found in \citetalias{Smith2020}. In these simulations a large proportion of the gas has been consumed by 1~Gyr, leaving isolated, extremely compact knots of dense gas within a low density ambient medium.
The gas discs in those simulations are extremely thin.
For the simulations that we do show in Fig.~\ref{fig_proj}, the \textit{IMFav} and \textit{IMFsam} simulations are
qualitatively similar. When SNe are the only source of feedback, the disc is thick and highly disordered. SNe superbubbles
blow transient holes in the disc, with star formation restricted to the remaining dense structures. When ionizing radiation
is the only source of feedback, the disc lacks the large holes and is more regular. There is a complex filamentary structure,
with dense star forming clouds embedded in lower density gas. The disc is not as thick as the \simSN{} simulations, but it
has not collapsed vertically as in the \simNoFB{} and \simPE{} simulations. The simulations with all feedback channels 
switched on are qualitatively similar to the \simPI{} simulations, particularly in the \textit{IMFav} case. However,
the \simSNPIPE{} \textit{IMFsam} simulation produces small, SN-driven cavities more regularly than the \textit{IMFav}
equivalent.

Fig~\ref{fig_proj_hii} shows the ratio between ionized and total hydrogen surface densities, face-on and edge-on. It is
therefore a form of projected ionization fraction. With reference to Fig.~\ref{fig_proj}, 
it can be seen that ionized regions
largely trace the diffuse gas. However, in the simulations with photoionization feedback there are small regions of high
ionization fraction embedded in dense gas. These are \ion{H}{ii} regions around ionizing sources. There is a qualitative
difference in the distribution and size of \ion{H}{ii} regions between the \textit{IMFav} and \textit{IMFsam}
simulations. The former simulations contain many small \ion{H}{ii} regions, speckled throughout the disc. The \textit{IMFsam}
simulations produce fewer, but larger \ion{H}{ii} regions. This difference is not a transient effect and persists
throughout the course of the simulations. 

The reason for this difference lies in the differing discretization of the ionizing sources
when IMF averages are used as opposed to explicitly tracking the emission from individual massive stars. With the
former approach, every star particle of the same age emits the same ionizing flux. However, the \textit{IMFsam} scheme
correctly ties the origin of ionizing photons to comparatively rare, bright sources. 
The total luminosity of ionizing
radiation per unit stellar mass is the same in both methods, by construction. However, the distribution of this luminosity
between sources makes a significant difference to the ability to photoionize gas. Despite having approximately half the
average global SFR, the \textit{IMFav} simulations with photoionization feedback keep roughly the same mass of gas
photoionized as the \textit{IMFsam} simulations. This is illustrated in Fig.~\ref{fig_ionmassloading} where we plot the
ratio between the photoionized mass (tagged as belonging to an \ion{H}{ii} region by our sub-grid model) and the mass
in star particles younger than 10~Myr. The pale lines show the instantaneous value at every output time (5~Myr) but as
this is extremely noisy we also plot a 50~Myr moving average. Fig.~\ref{fig_ionmassloading} demonstrates that
stars are almost twice as efficient at creating \ion{H}{ii} regions when the IMF averaged luminosities are assigned to
star particles.

The mass of gas in an \ion{H}{ii} region is determined by the balance between photoionization and recombination.
It is therefore linearly proportional to the ionizing photon rate and inversely proportional to the square of the
ambient density.
The net ionizing photon rate per unit stellar mass is the same regardless of the discretization method adopted, so
Fig.~\ref{fig_ionmassloading} implies that ionizing sources are typically embedded in moderately denser gas in the \textit{IMFsam}
case. Indeed, Fig.~\ref{fig_sfsn_pdfs} indicates this to be the case.
The ionizing photon budget is dominated by rare, massive
stars. With the IMF averaging approach, every star particle is an ionizing source. A $20\,\Msun$ star particle
will initially produce ionizing photons at a rate of approximately $10^{48}\,\mathrm{s}^{-1}$, this rate only
beginning to decrease after approximately 3~Myr, corresponding to the lifetime of the most massive stars.
An ionizing photon rate
of $10^{48}\,\mathrm{s}^{-1}$ corresponds to a roughly $16\,\Msun$ star. 
Once even a small quantity of gas is photoionized, the resulting D-type expansion allows the source to grow an \ion{H}{ii} region.
It will be able to photoionize
a $20\,\Msun$ gas cell as long the gas density is less dense than $216\,\mathrm{cm^{-3}}$. 
We find that for the
\textit{IMFav} \simSNPIPE{} simulation 53\% of the total ionizing luminosity is emitted by sources for which the resulting \ion{H}{ii} region
is resolvable by at least one cell at birth. Despite half the ionizing photons being `wasted' by under-resolution, this feedback is
still highly efficient.
Because \textit{every} star particle immediately begins emitting radiation upon creation, as soon as a
star forming region creates a single star particle the further collapse of the cloud can begin to be arrested. This effect
leads to the shifting of the birth density PDF towards the SF threshold in Fig.~\ref{fig_sfsn_pdfs}. 

By contrast, even
though explicit IMF sampling leads to far brighter sources they are significantly rarer. Only $8\%$ of $20\,\Msun$ star particles are assigned at least one star more massive than $16\,\Msun$ when the \textit{IMFsam} scheme is used. This means that on
average a proportionally larger amount of stellar mass has to be created before an ionizing source capable of affecting
the star forming cloud is born, resulting in a higher effective cloud-scale star formation efficiency. This has the added effect that star forming clumps can collapse to higher densities, as seen
in Fig.~\ref{fig_sfsn_pdfs}. 
This means that the sources are more likely to be born in higher density gas, resulting in a lower total \ion{H}{ii}
region mass for the same net ionizing luminosity per unit stellar mass. It is important to note that 
despite being embedded in typically higher density gas, the higher luminosity sources produced by
the \textit{IMFsam} approach means that 85\% of the total ionizing luminosity is associated with resolved \ion{H}{ii} regions
(by one cell) at birth, compared to 53\% (i.e. the opposite trend to what would be expected if resolution effects were
driving the difference between the approaches).
In summary, while the \textit{IMFav} simulations produce
many small \ion{H}{ii} regions around low brightness sources, the \textit{IMFsam} simulations produce fewer, but larger,
\ion{H}{ii} regions around massive stars.
For the reasons explained above, the latter scenario is less
efficient at regulating star formation.
\vspace{-4ex}
\section{Discussion} \label{Discussion}
We now discuss our findings in greater detail and place them in the context of some other relevant works. In Section~\ref{SN sampling}
we review best practices for schemes that trigger discrete SNe from IMF averaged rates. We show that if SNe are the only
form of feedback, this will yield identical results to explicit IMF sampling, which is what we see in our simulations.
In Section~\ref{PI sampling} we discuss
our contrasting result that the photoionization feedback \textit{is} in fact sensitive to the choice of method. We consider what these
findings might mean for other non-SN feedback channels and how this sensitivity depends on resolution. We also explore whether a simpler
toy model can capture the effects produced by explicit IMF sampling.
\vspace{-4ex}
\subsection{Insensitivity of supernova feedback to IMF averaging vs IMF sampling} \label{SN sampling}
\subsubsection{Best practices for discretizing IMF averaged SN rates}
\cite{MacLow1988} showed that as long as the number of SNe a star particle will produce over its lifetime is more than $\sim10$ then
the injection of SN feedback can be modelled as a continuous injection of energy. However, at higher resolution resolving the effects
of discrete SNe becomes important \citep[see e.g.][]{Su2018,Applebaum2020}. \cite{Keller2020} also demonstrate that simply injecting
the entire feedback budget of the star particle in one event also deviates from simulations that account for a spread of SN events
as a function of time. The impact of SN feedback is highly dependent on the clustering properties of the SNe 
(\citealt{Sharma2014,Yadav2017,Gentry2017,Gentry2019,Fielding2017,Fielding2018,ElBadry2019} and \citetalias{Smith2020}).
It is therefore important that if explicit IMF sampling is not used SNe must still be modelled as individual
events sampled from the IMF averaged SN rates (if the resolution is high enough to detect the effects).

After an expected number of SNe, $\bar{N}_\mathrm{SN}$, that will occur in a given time-step has been determined from the SN rate
(in general a function of star particle age and metallicity), \textit{initial} particle mass and time-step size (as described in 
Section~\ref{Sim methods}), there are two main approaches to converting these rates into discrete SN events. In the first approach,
Bernoulli trials can be carried out every time-step to determine whether a SN occurs. This essentially represents the flipping of a 
biased coin. If $\bar{N}_\mathrm{SN} \leq 1$, then a Bernoulli trial is carried out with a probability of success equal to 
$\bar{N}_\mathrm{SN}$ to decide whether a single SN occurs. If $\bar{N}_\mathrm{SN} > 1$, then the number of SNe that occur is 
at least equal to
the integer part of $\bar{N}_\mathrm{SN}$ with a Bernoulli trial carried out with a probability equal to the fractional part of 
$\bar{N}_\mathrm{SN}$ to determine whether an additional SN is added (see e.g. \citealt{Stinson2010} and the RIMFS scheme of
\citealt{Revaz2016}). This approach enforces a relatively smooth sampling of the rates, implicitly assuming that SN events are
spread out evenly through whatever interval was used to determine the SN rates in the first place. When star particle masses
are small, it also implicitly assumes that massive stars are distributed evenly across the particles which is essentially equivalent
to assuming that the IMF is well sampled in each particle. In other words, these schemes capture the mean SN rates but do not
capture the noise that emerges when considering a truly random sampling of an IMF in a single star particle.\footnote{Which may
or may not be desirable, depending on the manner in which one believes the IMF to be sampled in nature. Nonetheless, it is inconsistent
with the form of explicit IMF sampling presented in the majority of this work.}

Sampling from the Poisson distribution captures this noise. The probability of $N_\mathrm{SN}$ occurring given $\bar{N}_\mathrm{SN}$ is
\begin{equation}
\mathrm{P_\mathrm{Pois}}\left(k = N_\mathrm{SN} \right) = \frac{\bar{N}_\mathrm{SN}^k e^{-\bar{N}_\mathrm{SN}}}{k!}.
\end{equation}
Sampling from this distribution will produce a number of SN events that are scattered about $\bar{N}_\mathrm{SN}$, capturing the
noise, in contrast to the Bernoulli trial method. 

It is worth pointing out that there is a great deal of confusion in the literature
between conducting Bernoulli trials and sampling the binomial distribution, with the latter frequently being used when referring to the former.\footnote{This is because Bernoulli trials
are sometimes referred to as binomial trials (as distinct from, but related to, the binomial distribution).
Anecdotal evidence gathered by
tracing such references back to the details of published methods papers or publicly available code, where available, suggests that references to sampling from a binomial distribution most likely always refer to a Bernoulli trial scheme. For clarity, we will consistently refer to Bernoulli trials
in this work, even when citing authors that prefer the alternate phrasing.} 
The binomial distribution describes the probability of achieving a given number of successes from a finite number of independent Bernoulli trials that have the same probability of success. In the limit that the number
of trials is one, the binomial distribution converges to the Bernoulli distribution, but this is a trivial statement. Alternatively,
if multiple time-steps all have the same SN rate then the total number of SN produced across all the time-steps 
will represent a draw from the binomial
distribution. But one cannot ``sample from the binomial distribution'' to produce a number of SNe given an expected number of
SNe for an individual star particle.
This confusion also frequently leads to the statement that Bernoulli trial and Poisson sampling schemes are generally equivalent, 
because the binomial and Poisson distributions converge in the limit of many trials with low probability of success.
This is true across many time-steps if $\bar{N}_\mathrm{SN}$ is sufficiently small \textit{and} the SN rate does not change between
each time-step. However,
within each of the small time-steps the Bernoulli trial and the Poisson draw are only equivalent when $\bar{N}_\mathrm{SN}$ is 
very small such that
$\mathrm{P_{Pois}}\left(N_\mathrm{SN} > 1\right) \rightarrow 0$, which is a subtly different convergence criterion. Thus, both schemes
may produce the same number of SNe averaged over some long timescale, but the temporal clustering will be different, unless this
criteria is met.
In practice, this limit is frequently desirable because in order to
correctly capture the effects of SN clustering we wish to treat SNe individually. However, it is still more formally correct to
draw from a Poisson distribution (technically allowing for the possibility that more than one SNe occur in a given time-step) but impose a time-step limiter
to ensure that $\bar{N}_\mathrm{SN}$ is very small such that it is extremely unlikely that more than one SNe occurs. That is the approach
we take in this work.
\vspace{-4ex}
\subsubsection{The equivalence of discretized SN rates and explicit IMF sampling}
\begin{figure}
\centering
\includegraphics{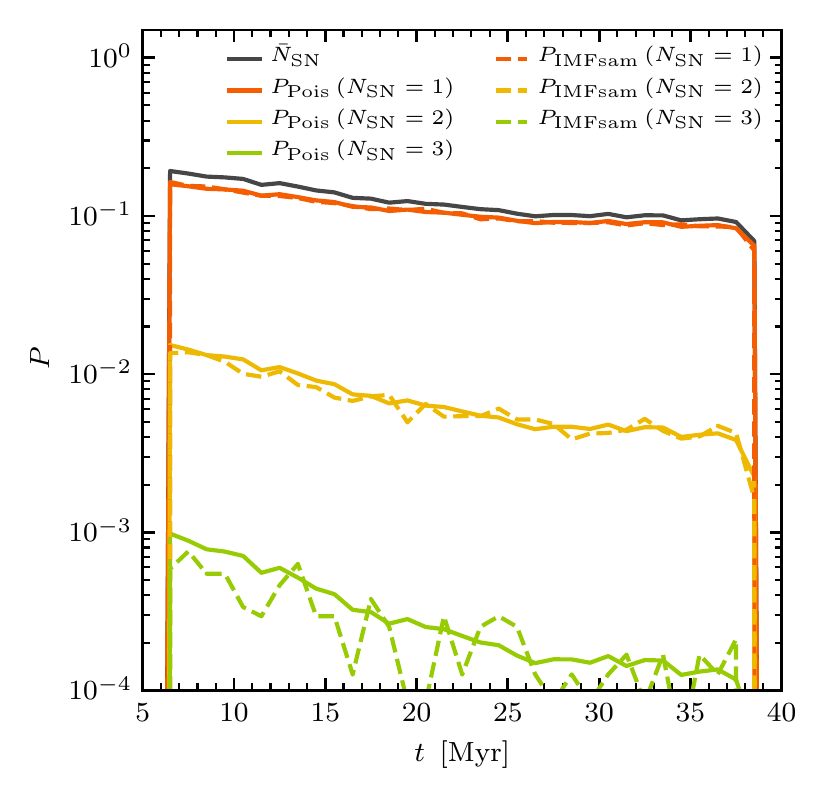}
\vspace{-6ex}
\caption{We demonstrate that for a $420\,\Msun$ star particle with a time-step of 1~Myr \citep[as used in][]{Applebaum2020}, the timing
of SNe are well described by Poisson distributions. For each time-step in the life of the star particle, we plot the expected number
of SNe, $\bar{N}_\mathrm{SN}$, in black. This is also equivalent to the probability of triggering a single SNe when a simple Bernoulli trial
scheme is used. The solid coloured lines show the probability of one, two or three SNe being produced in the time-step assuming a Poisson
distribution with a mean of $\bar{N}_\mathrm{SN}$. 
The dashed coloured lines show the equivalent probability derived from our Monte Carlo experiment, 
obtained as the fraction of particles that produce one, two or 
three SNe in the time-step when we explicitly populate particles with stellar masses drawn from the IMF. The total mass of particles is
$10^7\,\Msun$. It can be seen that they follow Poisson distributions (although the $N_\mathrm{SN}=3$ case is noisy due to the rarity
of that number of SNe being produced).
\vspace{-4ex}}
\label{fig_sn_prob} 
\end{figure}
\cite{Applebaum2020} use simulations of an isolated $M_\mathrm{vir}=10^9\,\Msun$ dwarf to directly compare an explicit IMF sampling
feedback scheme to an IMF averaged scheme similar to ours. 
Having established elsewhere in the work (with cosmological zoom-in simulations) that continuous injection of SN energy results in
a weaker impact of feedback, they discretize their SNe by sampling from the IMF averaged rates (which they refer to as 
`quantized feedback'). The feedback used is comprised of Type II SNe, H$_2$ dissociating Lyman-Werner (LW) radiation, Type Ia SNe and
stellar winds, although the latter two channels remain IMF averaged in all simulations. 
They find that the quantized feedback simulations result in 10\% more star formation and 30\% more cold ($T<1000\,\mathrm{K}$) gas compared to simulations that use explicit IMF sampling. 
They posit two possible causes for this. The first is that the difference is being
driven by the different distributions of LW sources between the two schemes. They suggest that this difference occurs because the SNe
and dominant LW sources are co-spatial when explicit IMF sampling is used, but it could also be directly related to the presence of
rarer, brighter sources with this method. The second suggestion is that sampling SNe from IMF averaged rates could produce a different
clustering of SNe than the explicit sampling scheme. As we have already argued, this should not be the case if the schemes are constructed in
a consistent manner. They carry
out a Monte Carlo experiment to compare the timing of SNe produced from star particles that are populated with stars from the IMF as
opposed to those that sample the IMF averaged SN rates. With their adopted particle mass of $420\,\Msun$ and time-step of 1~Myr, they find that
approximately 25\% of particles will experience at least one time-step in which more than one SN occurs when the explicit IMF sampling
is used. When they use Bernoulli trials to sample the rates this never happens (because $\bar{N}_\mathrm{SN}$ is always less than one).
\cite{Applebaum2020} note that
they could possibly have seen different results if they had adopted Poisson sampling instead of their Bernoulli trial scheme.

When we repeat their experiment with Poisson sampling instead of Bernoulli trials, we find that the fraction of particles experiencing
multiple SNe in a time-step is identical to the explicitly sampled IMF particles, to within 0.3\%. Furthermore, we demonstrate in
Fig.~\ref{fig_sn_prob} that the SNe rates that emerge from the particles that are populated from the IMF are described very well
by Poisson distributions. For a given 1~Myr time-step, we plot the expected number of SNe, $\bar{N}_\mathrm{SN}$, and the probability
that a particle will produce one, two or three SNe in that time-step, assuming a Poisson distribution with a mean of 
$\bar{N}_\mathrm{SN}$. We over-plot (in dashed lines) the fraction of star particles from our Monte Carlo experiment (which total
$10^7\,\Msun$) that produce that number of SNe in that time-step. It can be seen that the two agree very well (although the
$N_\mathrm{SN}=3$ fractions are very noisy because of the rarity of such an event). Indeed, we find that the number of SNe in a time-step are
well described by Poisson distributions whatever the choice of particle mass and time-step, in contrast to a Bernoulli trial scheme.

With star particles of mass $420\,\Msun$, a time-step
of 1~Myr is too large for the Bernoulli trial approach to converge with the more consistent Poisson distribution. 
However, the discrepancy still
appears to be relatively small. It is possible that the use of a `blast-wave feedback' model \citep{Stinson2006} exacerbates the effect
since the adopted sub-grid evolution of the SN remnant (which affects the length of time for which cooling is shut off in this model)
is different for a single injection of the energy of multiple SNe compared to spreading those events into multiple independent
injections. Regardless, if the SN clustering is the primary cause of the difference between the quantized feedback
and explicitly sampled IMF schemes, then it results from an inconsistent implementation of the SN rate sampling. We suspect that the
difference is more likely to be caused by the first explanation (differences in the distribution of LW sources). Whatever the cause,
we feel that the results of \cite{Applebaum2020} should only be interpreted as reinforcing the need for explicit IMF sampling when
pre-SN feedback is included. They do not demonstrate that the stochastic sampling of IMF averaged SN rates is an inferior approach
to explicit IMF sampling \textit{in the absence} of pre-SN feedback channels.

The fact that the distribution and timing of SN events among star particles is identical when we use
quantized feedback or explicit IMF sampling also shows that the stellar mass implied by the SN events is consistent with the dynamical
mass to the same degree in both cases. Both quantized sampling of SN rates and explicit IMF sampling allow the fraction of star particle mass occupied by low mass stars to vary, by construction. Some particles will have more massive stars per unit stellar mass than the IMF average,
while others have correspondingly fewer. However, in the case of quantized sampling of SN rates, the inventory of the star particle is not
known until a SN occurs, indicating that a SN progenitor implicitly existed in the particle from its birth. This means that it is
impossible for the level of non-SN feedback (which will also be pre-SN)
generated by the star particle to made consistent with the stellar inventory that is `discovered'
once SNe have occurred. This can be thought of as an inconsistency between the implied 
distributions of stellar masses responsible for the SN and non-SN feedback
\citep[see][fig. 3 for an illustration of this effect]{Applebaum2020}. With explicit IMF sampling, the stellar inventory is known from the
birth of the particle, so the level of pre-SN feedback can be made consistent with the number and timing of the SNe produced by the particle.
In the absence of non-SN feedback, quantized SN feedback and explicit IMF sampling are equivalent.

To summarise this section, we find that if SNe are the only source of stellar feedback, triggering discrete, individual SNe
from IMF averaged rates gives identical results to explicit sampling of the IMF at the moment of a star particle's creation
(assuming the sampling can be carried out without biasing the IMF, as described in Section~\ref{pop IMF methods}). We
caution that this is only the case when the time-step in which the sampling is carried out is sufficiently small and that, in general,
sampling from a Poisson distribution is the correct procedure. We have demonstrated this with our simulations
shown in Section~\ref{results} but have described in this section why this must necessarily be true for any consistently constructed
scheme. Therefore, if SNe are the only form of stellar feedback considered in a simulation nothing is gained by adopting an explicit
IMF sampling scheme, which carries with it a penalty in terms of code complexity and (potentially) memory requirements.

\subsection{Sensitivity of non-supernova feedback to IMF sampling} \label{PI sampling}
\subsubsection{Under what circumstances does the sensitivity arise?}
\begin{figure}
\centering
\includegraphics{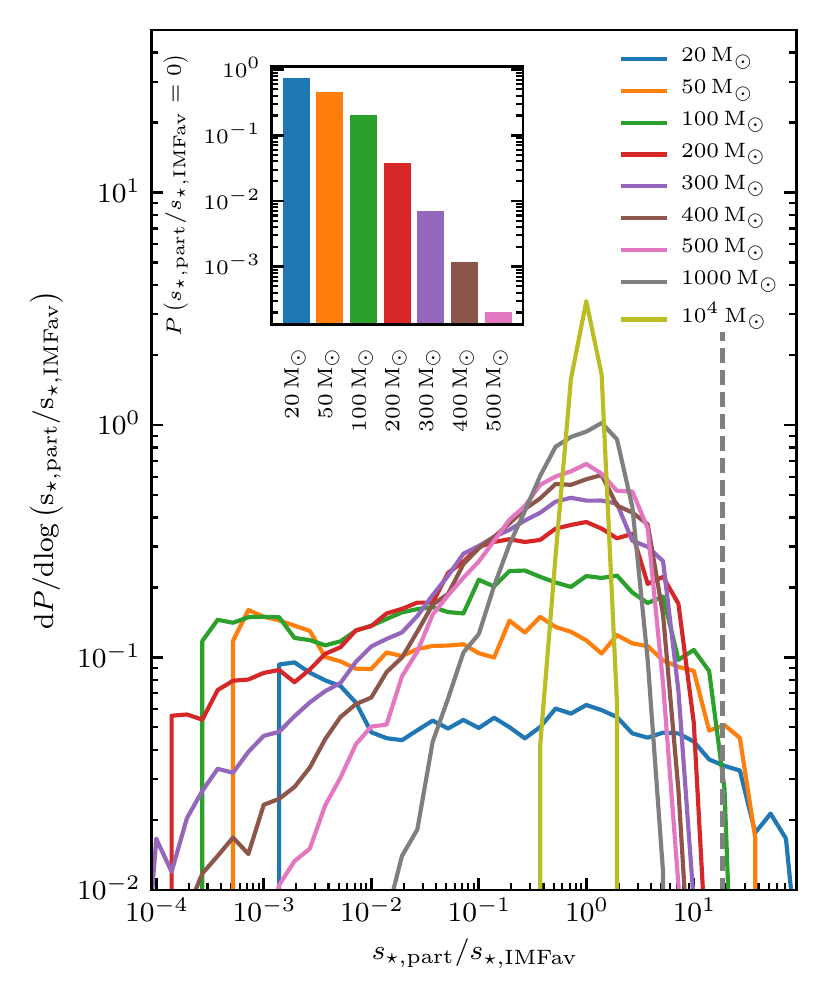}
\vspace{-6ex}
\caption{The distribution of the specific ionizing photon rate, normalized by the IMF average, for $10^7\,\Msun$ of stellar mass
divided into star particles and populated from the IMF using the adjusted target scheme. 
The lines show PDFs for different particle masses for those particles that have
non-zero ionizing luminosities, while the inset plot shows the fraction of particles that have not been assigned a star that produces
ionizing photons. The smaller the particle mass, the larger the spread about the IMF averaged value. The vertical dashed line indicates
the maximum physical value for our adopted stellar mass range, corresponding to a $100\,\Msun$ star. High resolution particles can exceed
this specific ionizing photon rate if they have been assigned a more massive star than their dynamical mass, but as discussed in the 
main text, this does not result in over-bright sources.\vspace{-4ex}}
\label{fig_sample_rad} 
\end{figure}
SN feedback is unique among stellar feedback channels in that it is composed of discrete, instantaneous (relative to other
astrophysical timescales) events. All other forms of stellar feedback (e.g. radiation or stellar winds) are more continuous in nature,
coupling to gas over an extended period of time. The degree to which the impact of the feedback is sensitive to the details of 
IMF sampling depends on the resolution in two ways. Firstly, does explicit IMF sampling produce significant enough inhomogeneities
from particle to particle for results to deviate from an IMF averaged approach? The larger the star particle mass, the smaller
the spread in feedback properties. Secondly, is the gas resolution high enough that the difference causes a resolvable effect?
This also depends on other simulation details, such as the star formation prescription.

In the simulations we presented in this work, the predominant difference between the \textit{IMFav} and \textit{IMFsam} simulations
was the efficiency of the photoionization feedback. We can study how the level of inhomogeneities in the ionizing photon rate from 
particle to particle varies as a function of particle mass by carrying out another idealized Monte Carlo experiment. We again divide
a stellar mass budget of $10^7\,\Msun$ into particles of mass $\mpart$. We populate these with stellar masses drawn from
the IMF using the adjusted target scheme. We can then obtain the ionizing photon rate, $S_{\star,\mathrm{part}}$, 
for the particles at $t_\mathrm{part}=0$ by summing
the contributions from their component stars, as is done in the simulation. To enable comparison between different particle masses,
we then obtain the specific ionizing photon rate, $s_{\star,\mathrm{part}}=S_{\star,\mathrm{part}}/\mpart$, for each particle.
Fig.~\ref{fig_sample_rad} shows PDFs of $s_{\star,\mathrm{part}}$ for the ensemble of particles at a given $\mpart$, normalized
by the IMF averaged specific ionizing photon rate, $s_{\star,\mathrm{IMFav}}$. Therefore, the spread around unity indicates the level
of inhomogeneities amongst the population of star particles compared to using the same IMF averaged ionizing photon rate for all particles.
It is possible for a star particle to have an ionizing luminosity of zero because it has not 
been assigned any stars more massive than $7\,\Msun$. Below this mass we assume the star emits no ionizing photons for simplicity.
The ionizing luminosity is already negligible at this mass so this is not an issue, but this does result in a discrete feature
at $s_{\star,\mathrm{part}}=0$ which cannot be meaningfully represented in our logarithmic PDFs. We therefore include an inset figure
that shows the probability that a particle has an ionizing luminosity of zero. Thus the integral 
under the PDF and the value of corresponding bar in the inset plot sum to unity together.

As expected, Fig.~\ref{fig_sample_rad} shows that the spread around the IMF averaged specific luminosity reduces as the particle mass
increases. This is because the IMF is more completely sampled within the particle for a larger particle mass i.e. 
$s_{\star,\mathrm{part}}$ converges with $s_{\star,\mathrm{IMFav}}$ as $\mpart$ goes to infinity. It is important to note that this does
not mean that inhomogeneities are not important, rather that a simulation with a large particle mass is incapable of resolving them.
If the particle mass is large enough that inhomogeneities vanish, then using IMF averaged values will not result in any further loss of
fidelity in the simulation and so this simpler approach can be adopted. With $\mpart=10^4\,\Msun$, the largest particle mass we consider,
each particle samples the IMF relatively well, so $s_{\star,\mathrm{part}}$ is very tightly distributed about the IMF average, deviating
by a factor of 2 at most. However, as the particle mass decreases, the spread about the IMF averaged value is no longer insignificant.
It already begins to span a factor of a few when $\mpart=10^3\,\Msun$. The distribution continues to become broader and flatter as the
particle mass as further decreased. A sharp cutoff is apparent at the low end of the PDF. As mentioned before, this originates from
particles that are assigned no ionizing sources. The smallest non-zero $s_{\star,\mathrm{part}}$ corresponds to the emission from
a single $7\,\Msun$ in the particle, so this cutoff occurs at higher values of $s_{\star,\mathrm{part}}$ for lower $\mpart$.
Correspondingly, the fraction of particles that produce no ionizing radiation (the inset panel) increases with decreasing $\mpart$, such
that when $\mpart=20\,\Msun$ only approximately a quarter of particles produce radiation. However, even if particles do emit radiation,
when $s_{\star,\mathrm{part}}$ is more than factor of a few below the IMF averaged value this emission is negligible.

The top end of the PDFs represent the rare bright sources. It can be seen that with $\mpart=300\,\Msun$, some particles can
have a specific ionizing luminosity an order of magnitude larger than the IMF averaged value. The maximum physically obtainable
specific ionizing luminosity for the stellar mass range we consider corresponds to emission from a $100\,\Msun$ star. This is 18.47 times
the IMF averaged value.\footnote{We confirmed in test simulations for \citetalias{Smith2020} that our results are relatively
insensitive to dropping this maximum stellar mass to $50\,\Msun$, which corresponds to a specific ionizing luminosity relative to the IMF average of 9.65.} This limit is indicated with a dashed line in Fig.~\ref{fig_sample_rad}. Low mass particles can exceed this limit when they
are assigned more stellar mass than their dynamical mass (a natural consequence of our IMF sampling scheme which is not particularly 
problematic, as discussed in Section~\ref{pop IMF}). Note that this does not result in an unphysically bright source in the simulation or
affect the clustering of bright sources. It simply indicates that a more physically correct determination of their specific ionizing
luminosity should account for the particles that have a mass discrepancy in the opposite sense, created by the adjusted target IMF sampling
scheme to ensure mass consistency over multiple particles.

Fig.~\ref{fig_sample_rad} demonstrates quantitatively how the ability to resolve inhomogeneities in the 
strength of ionizing sources varies as a function of particle mass. Similar scalings will result from any feedback source and are
relatively easy to determine. However, the impact of the inhomogeneities on the outcome of the simulation are harder to predict. 
By averaging over the IMF, one gains more sources that have the IMF averaged feedback strength at the cost of losing the
strongest sources. In our
simulations, using IMF averaged ionizing photon rates results in more efficient feedback. This is because 
a significant number of sources
with the IMF averaged
rate are able to start forming resolvable \ion{H}{ii} regions and sources are typically present earlier in the life of a star
forming cloud, 
so the penalty incurred by losing the brightest sources is outweighed by
benefits of increasing the overall number of ionizing sources. This positive impact on the overall feedback strength of IMF averaging will
not necessarily occur in all scenarios. If the IMF averaged ionizing photon rate was not enough to form resolvable \ion{H}{ii} regions
(e.g. because of a different resolution, star formation prescription etc.) then the reverse could occur since the simulation would
contain no sources bright enough to have an impact. Note also that in simulations of an individual GMC at a \textit{higher} resolution than
us, \cite{Grudic2019} found the opposite trend (i.e. more discretization of ionizing sources leads to more effective feedback). Thus,
it is difficult to predict the impact of IMF averaging vs. sampling in a given simulation. However, we would suggest that as the latter is
arguably more physical it should in general be adopted if there will be significant particle to particle inhomogeneities. 

These arguments apply to all forms of non-SN stellar feedback, not just ionizing
radiation. We possibly see the reverse trend affecting our photoelectric heating feedback in our simulations than affects photoionization
feedback. Photoelectric
heating is very ineffective in our simulations of dwarf galaxies, as can be seen in Fig~\ref{fig_sfr}. This is because 
of the low dust-to-gas ratio \citepalias[as discussed in][]{Smith2020}. The simulations with photoelectric heating alone
have similar SFRs to the simulation without any feedback. Nonetheless, the peak SFR in the \textit{IMFsam} 
version is marginally lower than the
\textit{IMFav} equivalent, possibly indicating that IMF averaging weakens the photoelectric heating feedback, in contrast to the
photoionization feedback which is strengthened. This could occur if only the brightest sources were able to cause sufficient heating to
have an impact and that spreading their luminosity among multiple sources (via IMF averaging) renders heating ineffective. However,
we caution that the difference is very slight and these galaxies are already in an unphysical regime, having fragmented into very dense
clumps and reached a very high SFR due to ineffective feedback.

\subsubsection{Can a simplified model be used to replicate the effects of explicit IMF sampling?}
\begin{figure}
\centering
\includegraphics{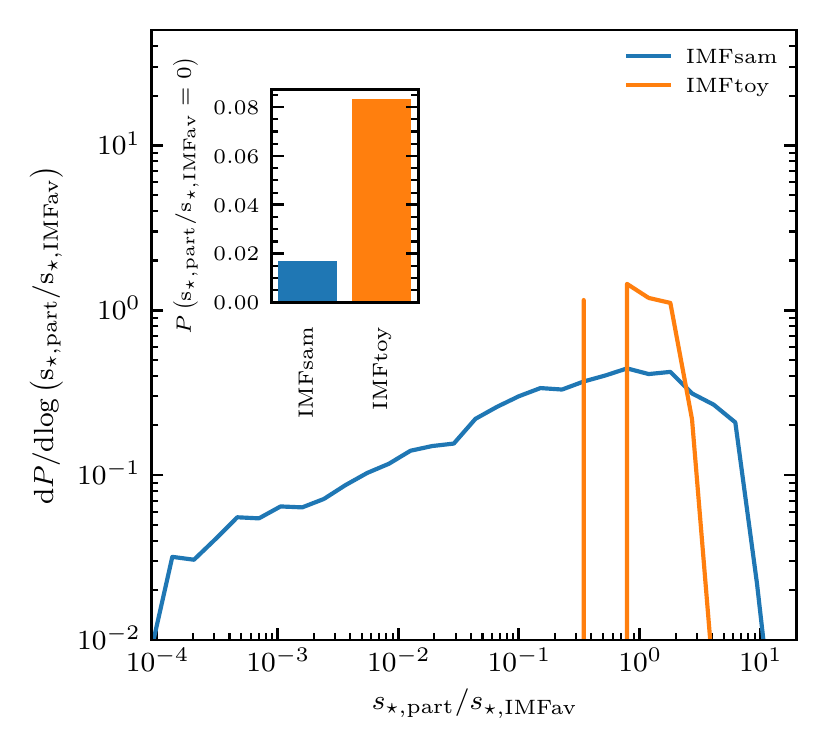}
\caption{Similar to Fig.~\ref{fig_sample_rad}, the distribution of the specific ionizing luminosity, normalized by the IMF average, for
a $10^7\,\Msun$ population of star particles. In this plot we show the results for $250\,\Msun$ particles with their luminosity obtained
by explicitly populating the particles from the IMF (\textit{IMFsam}) or by using the toy model of \protect\cite{Su2018} to modulate
the IMF averaged rates (\textit{IMFtoy}). The toy model results in a much narrower distribution than that which arises from sampling
the IMF directly because it does not account for the substantial variation in the luminosity of massive stars as a function of mass.
Note that the toy model forces the particles to adopt one of a discrete set of luminosities, so plotting this as a PDF is
sub-optimal but we do so to enable easy comparison to Fig.~\ref{fig_sample_rad}.\vspace{-3ex}}
\label{fig_sample_rad_ksu}
\end{figure}
\cite{Su2018} examines how discretizing stellar feedback affects dwarf galaxies, in this case with cosmological zoom-in simulations
with a mass resolution of $250\,\Msun$. In common with the other works previously mentioned, they find that modelling SNe as a
continuous injection of feedback, rather than as discrete events, substantially weakens its impact. Yet, in contrast to us,
they find that the efficiency of continuous non-SN feedback mechanisms (e.g. radiation and stellar winds) are largely unaffected
by accounting for the effects of IMF sampling as opposed to their default IMF averaged rates. This could indicate that the impact of 
the particle to particle inhomogeneities do not result in a resolvable effect at their mass resolution. However, they do not actually
perform explicit IMF sampling but instead attempt to replicate the effects via a toy model. At the point of creation, a star particle
is assigned a number of O stars, $N_\mathrm{O}$, drawn from a Poisson distribution with an expectation value of 
$\langle N_\mathrm{O} \rangle = m_\mathrm{part}/100\,\Msun$. Their feedback schemes then operate as in the fiducial case, but all
IMF averaged rates that are linked to massive stars (photoionization, photoelectric heating, UV radiation pressure, OB star winds and
core-collapse SN rates) are multiplied by a factor $N_\mathrm{O}/\langle N_\mathrm{O} \rangle$. When a SN occurs, $N_\mathrm{O}$ is 
reduced by one. It can therefore be seen that the strength of feedback will vary from star particle to particle according to the
number of assigned sub-grid O stars, but the net rates over multiple particles will maintain the IMF average. This scheme is much simpler
than explicitly sampling, assigning individual stars to particles and looking up their individual feedback budgets. The downside is that
the approximation does not completely capture the correct behaviour, for a number of reasons. 
As we have demonstrated in the previous section, stochastically
triggering SNe from IMF averaged SN rates already captures the correct clustering properties, as would be produced by explicit IMF
sampling.\footnote{As we have already discussed at length, this is true only if the sampling is carried out consistently. \cite{Su2018}
use Bernoulli trials. The combination of their
particle mass and typical time-step yields $\bar{N}_\mathrm{SN}\sim10^{-5}$, which is sufficiently small for Bernoulli trials
and Poisson sampling to converge.} Thus, boosting the IMF averaged SN rates will result in an additional over-enhancement of 
clustering. 

A potentially more problematic issue with the approximation is that it assumes massive stars are uniform in terms of their 
feedback budget, as the authors caution. A large amount of the scatter in the specific
ionizing luminosity of a star particle (shown in Fig.~\ref{fig_sample_rad}) originates not just from the variation in the number of
massive stars assigned to a star particle but also from the strong mass dependence of the ionizing luminosity produced by individual stars.
In other words, there is a significant variation in luminosity among OB stars.
Fig.~\ref{fig_sample_rad_ksu} shows the results of a similar Monte Carlo experiment, where we assign a $t_\mathrm{part}=0$
ionizing photon rate to 
$250\,\Msun$ star particles either by populating them with stars sampled from the IMF, as usual, or by using the toy model
of \cite{Su2018}. It can be seen that the toy model is not a good approximation to the correct distribution, with a much narrower
range of ionizing photon rates, producing far more stars close to the IMF average value and lacking the brightest sources. 

Additional
inconsistencies in the toy model arise as the star particles age. The multiplicative factor that modulates the 
feedback rates is reduced by $1/\langle N_\mathrm{O} \rangle$ when a SN occurs, 
regardless of when it occurs. An early SN indicates
a more massive progenitor, which in turn means a larger drop in the ionizing photon budget for the particle. This decrease in luminosity
as the most massive stars die is already accounted for in the time evolution of the IMF averaged rates. Another direct consequence of the
link between lifetime, stellar mass and photon budget is that the earliest SNe should occur in regions that have been exposed to the highest
ionizing flux, an effect not captured by the toy model. All of these issues taken together, but in particular that shown in 
Fig.~\ref{fig_sample_rad_ksu}, suggest that IMF sampling effects likely cannot be replicated by simply modulating the IMF averaged rates.
\vspace{-4ex}
\section{Summary and conclusions} \label{conclusion}
It is a common practice in simulations of galaxies to treat stellar mass in a homogenised manner, with star particles producing 
IMF averaged feedback. With the advent of simulations with increasingly higher baryonic mass resolution, we have examined the extent to
which this approximation is valid and under what circumstances it breaks down. We began by exploring methods of populating star
particles with inventories of stars drawn from the IMF. Because of the challenges of filling an arbitrary mass budget with a discrete
set of randomly drawn stellar masses, the goals of conserving mass (globally and locally) and faithfully reproducing the input IMF
are often in tension. We argued that because the IMF is almost never an emergent property because of resolution limits and missing physics,
an input IMF must be provided and reproduced accurately by the sampling scheme. 
This IMF may be fixed or it can vary based on some sub-grid
prescription, but it should not be biased by numerical issues such as the choice of star particle mass, since this will
affect the feedback budget. Sampling schemes must therefore prioritise reproduction of the IMF and the conservation of mass across
a population of star particles over ensuring that the mass of the stellar inventory assigned to the particle is consistent with the 
dynamical mass of the particle. Because exact N-body dynamics of individual stars are generally unresolved in galaxy formation simulations,
such a mass discrepancy is in general unimportant as long as an effort is made to minimize it. The discrepancy can also be resolved by
allowing additional mass transfer to/between star particles,
 but this brings its own complexities and issues. Instead, we consider simple schemes that
can easily be incorporated into existing star formation implementations without mass transfer. 
We carry out a series of Monte Carlo experiments to demonstrate
the properties of these schemes, with the following results:
\begin{itemize}
\item{Schemes that populate each star particle in isolation (i.e. without taking into account how other particles were populated)
inevitably bias the IMF. The stop before approach, which throws away the last drawn stellar mass (that which 
carries the total assigned mass
over the target), over-produces low mass stars and suppresses the high mass end of the IMF. The stop after approach, which always
keeps the last draw, reproduces the shape of the IMF but results in too high a normalization, over-producing stars of all masses.
The stop nearest approach, which retains a draw if keeping it results in a total assigned mass that is closer to the target, suffers from the same
biases as the stop before scheme, albeit to a lesser extent.}
\item{Because they fail to accurately reproduce the IMF, all of these schemes
result in an incorrect feedback budget. The relative size of this error diminishes with increasing particle mass, but is still
apparent when the particles have a mass of a few hundreds of $\Msun$.}
\item{We propose a scheme, which we refer to as adjusted target, that takes into account the degree to which previous particles overshot their target and selects a new target mass to compensate. Thus some star particles have more stellar mass assigned than their dynamical mass while the reverse is true for others, but the mass discrepancy is always minimized as much as possible. Our scheme results in the input IMF being perfectly reproduced for all star particle masses and produces the correct feedback budget 
(e.g. the number and timing of SNe, the
amount of ionizing radiation etc.).}
\end{itemize}

In the second part of this work, we carried out isolated simulations of an $M_\mathrm{vir}=10^{10}\,\Msun$ dwarf galaxy
with a baryonic mass
resolution of $20\,\Msun$. We included stellar feedback in the form of core-collapse SNe, photoionizing radiation 
and photoelectric heating, treated with the new sub-grid models described in \citetalias{Smith2020} and implemented in the
\textsc{Arepo} code. We compared two sets of simulations that either used IMF averaged rates for feedback or 
explicitly populated particles with discrete stars (using the adjusted target method). Our key findings are as follows:

\begin{itemize}
\item{If SNe are the only source of stellar feedback, triggering individual SNe via Poisson sampling of IMF averaged rates yields
identical results to explicitly sampling stellar masses from the IMF. This is because the distribution of the SNe in space and
time is essentially perfectly reproduced by a stochastic sampling of the rates as long as the time resolution is sufficiently high.}
\item{The impact of ionizing radiation is overestimated in our simulations when IMF averaged rates are adopted, particularly its ability to regulate SFRs. Approximately twice the mass of gas is photoionized per unit stellar mass when the IMF averaged rates are used
compared to the explicit IMF sampling equivalent simulations
because sources are typically embedded in lower density gas.
When IMF averaging is used every star particle immediately emits ionizing photons,
meaning that
clumps of gas begin to have their collapse disrupted as soon as star formation begins. This results in the production of many
small \ion{H}{ii} regions.
By contrast, IMF sampling correctly produces brighter but, 
crucially, rarer sources. 
This discretization means that on average a larger mass of stars must be formed before a significant
ionizing source appears. It also means that these sources are typically formed later in the evolution of a star forming clump and
are thus embedded in higher density gas. This results in less efficient feedback
and fewer but larger \ion{H}{ii} regions. It is important to note that this trend is \textit{not} driven by an inability to resolve
\ion{H}{ii} regions (see Section~\ref{results} for details).}

\item{The scenario we see in these simulations is not necessarily ubiquitous. The degree to which the
strength of IMF averaged feedback deviates
from the explicitly sampled case depends on a number of factors. Firstly, the star particle must be of a sufficiently small mass such
that appreciable inhomogeneities are apparent across the population of star particles. This will vary between different feedback
channels. Secondly, the impact of these inhomogeneities must itself be resolvable in the simulation. IMF averaging may also 
potentially decrease the effectiveness of feedback under certain conditions. Feedback strength goes up in our simulations when we use IMF
averaging because the resultant ionizing luminosity remains high enough to have a resolvable impact. We thus increase the number of
effective sources of feedback, despite the penalty of losing the strongest sources. However, if a feedback channel is completely
dependent on extremely rare, strong sources (e.g. the most massive O stars), then IMF averaging may result in a net reduction
in the feedback strength.}
\end{itemize}
\vspace{-2ex}
Given the complex dependence on resolution, the details of sub-grid models and
the highly non-linear behaviour of stellar feedback, it is difficult to predict a priori whether there will be a difference between
an IMF averaged feedback scheme and an IMF sampled scheme in any given simulation. However, the IMF sampling approach is 
always the more physically motivated approach. In other words, the best an IMF averaged scheme can achieve is to give results
that converge with an IMF sampling approach. Even then, it will only converge when the effects of sampling are unresolvable and thus
no more fidelity can be gained by explicitly populating star particles with discrete stars. Therefore, rather than trying to estimate
how much of an effect choosing IMF averaging over sampling will have on the resolved feedback a priori, we suggest that the choice
should simply be based on the level of inhomogeneities between particles for the chosen particle mass. This varies between different
feedback channels, but based on the variation in the specific ionizing luminosity between particles (see Fig.~\ref{fig_sample_rad}), 
we very conservatively suggest that explicit IMF sampling should be used when the particle mass is less than $\sim500\,\Msun$.

In this work, we have used a single, fixed IMF \citep{Kroupa2001} in our tests. However, as we discussed, our techniques and findings are
applicable to any IMF, including those that can vary based on local properties. Variable IMFs can already be adopted by IMF averaged
feedback schemes by adding additional dimensions to lookup tables, but they can be treated in a very natural fashion by explicit sampling
schemes, allowing very fine grained control of stellar populations on-the-fly. This would be an interesting avenue of future research. We
have also only considered single stars. The sampling scheme could be extended to explicitly account for
binary systems. The inclusion of binary stellar evolution results in a modest increase in the total ionizing photon budget and extends the
production of ionizing radiation to later times \citep[see e.g.][]{Eldridge2009}, as well as introducing an additional population of late-time
core-collapse SNe \citep{Zapartas2017}. It would also allow feedback channels that depend on binary systems (e.g. Type Ia SNe, jets from High Mass X-ray Binaries and OB runaways
etc.)
to be treated in a self-consistent manner. Thus, using explicit IMF sampling in galaxy formation simulations not only allows us to
capture the inhomogeneous distribution of stellar feedback sources but may also provide a more direct link to models of stellar evolution
than IMF averaged approaches.
\vspace{-4ex}
\section*{Acknowledgements}
The author is grateful to Elaad Applebaum, John Forbes and Chia-Yu Hu for helpful discussions and to Rachel Somerville, Greg Bryan
and the anonymous reviewer for comments which
improved this manuscript.
The work of M.C.S. was supported by a grant from the 
Simons Foundation (CCA 668771, L.E.H.).
The simulations were run on the Flatiron Institute’s research computing facilities (Iron and Popeye compute clusters), supported by the Simons Foundation.
\vspace{-4ex}
\section*{Data Availability Statement}
The data underlying this article will be shared on reasonable request to the corresponding author.
\bibliographystyle{mnras} 
\bibliography{imf_paper_arxivv2}
\appendix
\section{Robustness to stochasticity} \label{stoch}
\begin{figure}
\centering
\includegraphics{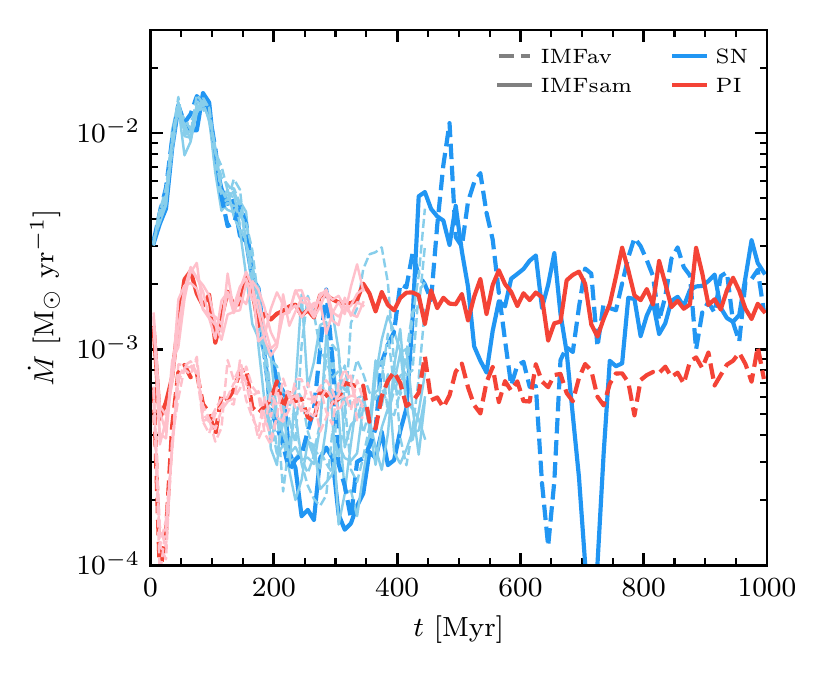}
\caption{SFRs for our perturbed re-simulations (in the lighter shades) with the fiducial simulations for comparison (bold colours).
The \textit{IMFav} and \textit{IMFsam} \simSN{} simulations are consistent with each other within the magnitude of the scatter between
re-simulations. By contrast, the differences between the \textit{IMFav} and \textit{IMFsam} \simPI{} simulations are much
larger than the scatter within a set of re-simulations.}
\label{fig_sfr_repeat} 
\end{figure}
Simulations of galaxy evolution are inherently chaotic to some degree, meaning that small perturbations introduced by, for example,
seemingly minor differences in initial conditions, choice of random number generator seed value, floating-point round-off and
non-deterministic behaviour of parallelised codes can lead
to measurable large scale differences in the outcome \citep{Keller2019,Genel2019}. In order to approximately assess how the differences
between our various feedback schemes compare to the magnitude of this stochastic uncertainty, we perform some re-simulations of the
early stages of the \simSN{} and \simPI{} simulations with both the \textit{IMFav} and \textit{IMFsam} schemes. For each fiducial
simulation, we carry out four additional re-simulations. In each re-simulation we use different seed values for random number generators.
We also perturb the position of every gas cell, star particle and dark matter particle in the initial conditions by moving it 0.1~pc in
a random direction. The SFRs for these simulations can be seen in Fig.~\ref{fig_sfr_repeat}, with the fiducial simulations shown
in the bold colours and the re-simulations shown in a lighter shade.

We were limited by computational expense from completely re-simulating the full 1~Gyr of our fiducial simulations, but it can
be seen that for the 450~Myr re-simulated the \textit{IMFav} and \textit{IMFsam} \simSN{} simulations are consistent with each other
within the range of stochastic scatter. We are therefore confident in our assertion that IMF averaging and IMF sampling give essentially
identical results if SNe are the only source of feedback. The \simPI{} re-simulations have a very tight scatter about the fiducial
simulations, such that the magnitude of the stochastic uncertainty is much smaller than the difference between the \textit{IMFav} 
and \textit{IMFsam} schemes.
\end{document}